\documentclass[final]{IEEEtran}
\pdfoutput=1
\usepackage{amsthm,amssymb,graphicx,multirow,amsmath,color,amsfonts}
\usepackage[update,prepend]{epstopdf}
\usepackage[noadjust]{cite}
\usepackage[latin1]{inputenc}
\usepackage{tikz}
\usetikzlibrary{arrows,calc}		
\usepackage{bbm} 
\usepackage{pdfpages}
\usepackage{flushend}
\usepackage{tabulary}
\usepackage{multirow}
\usepackage{comment}

\def\nbx{{\mathbf{x}}}
\def\nby{{\mathbf{y}}}
\def\nbz{{\mathbf{z}}}
\def\nb0{{\mathbf{0}}}
\def\nb1{{\mathbf{1}}}

\def\nbA{{\mathbf{A}}}
\def\nbB{{\mathbf{B}}}

\def\nbH{{\mathbf{H}}}
\def\nbI{{\mathbf{I}}}

\def\nbQ{{\mathbf{Q}}}
\def\nbR{{\mathbf{R}}}

\def\ncalA{{\mathcal{A}}}

\def\ncalI{{\mathcal{I}}}

\def\nbbE{{\mathbb{E}}}

\def\nbbN{{\mathbb{N}}}

\def\nbbP{{\mathbb{P}}}

\def\nbbR{{\mathbb{R}}}

\newtheorem{lemma}{Lemma}

\newtheorem{theorem}{Theorem}

\newtheorem{cor}{Corollary}

\newtheorem{remark}{Remark}

\DeclareMathOperator{\Tr}{Tr}

\DeclareMathOperator{\Pois}{Pois}

\allowdisplaybreaks 
\begin{document}
\graphicspath{{./Figures/}}
\title{Wireless Backhaul Networks: Capacity Bound, Scalability Analysis and Design Guidelines
}
\author{
Harpreet S. Dhillon and Giuseppe Caire 
\thanks{This paper will be presented in part at IEEE ISIT, Honolulu, 2014~\cite{DhiCaiC2014a,DhiCaiC2014}.}
\thanks{The authors are with the Communication Sciences Institute (CSI), Department of Electrical Engineering, University of Southern California, Los Angeles, CA 
(email: \{hdhillon; caire\}@usc.edu).}
\thanks{
This work was supported by NSF CCF-1161801 and a gift from Intel for research on 5G mm-wave wireless networks. Revised: \today.}
}

\maketitle

\begin{abstract}
This paper studies the scalability of a wireless backhaul network modeled as a {\em random extended network} with multi-antenna base stations (BSs), where the number of antennas per BS is allowed to scale as a function  of the network size. The antenna scaling is justified by the current trend towards the use of higher carrier frequencies, which allows  to pack large number of antennas  in small form factors. The main goal is to study the per-BS antenna requirement that ensures scalability of this network, i.e., its ability to deliver non-vanishing rate to each source-destination pair. We first derive an information theoretic upper bound on the capacity of this network under a general propagation model, which provides a lower bound on the per-BS antenna requirement. Then, we characterize the scalability requirements for two competing strategies of interest: (i) {\em long hop}: each source-destination pair minimizes the number of hops by sacrificing multiplexing gain while achieving full beamforming (power) gain over each hop, and (ii) {\em short hop}: each source-destination pair communicates through a series of short hops, each achieving full multiplexing gain. While long hop may seem more intuitive in the context of massive multiple-input multiple-output (MIMO) transmission, we show that the short hop strategy is significantly more efficient in terms of per-BS antenna requirement  for throughput scalability. As a part of the proof, we construct a scalable short hop strategy and show that it does not violate any fundamental limits on the spatial degrees of freedom (DoFs).
\end{abstract}


\section{Introduction} \label{sec:intro}

To handle increasing wireless data traffic, cellular networks are undergoing a paradigm shift from a well-planned deployment of large tower-mounted and high power  base stations (BSs) to an organic capacity-driven deployment of smaller and lower power BSs, often called {\em small cells}~\cite{QuedelB2013}.  While conventional BSs are typically connected through a high capacity 
wired backhaul network, the same is not true for small cells, which may have to be densely deployed at more adverse locations~\cite{SCFM2013}. For such systems, the backhaul represents one of the major bottlenecks in terms of throughput and deployment cost. A more economically viable alternative is to have wireless backhaul. The general idea is to establish high capacity wireless links that can carry cellular data from a cell to another cell and/or to some gateway node connected to the backbone network, possibly through multiple hops~\cite{CamComM2014}. In this work, we focus on the design of throughput scalable wireless backhaul systems, disregarding the cellular access links (uplink/downlink in each cell), which are assumed to operate in a different frequency band. As such, each BS is at the same time a source of traffic (generated by the users in its cell), a destination of traffic (to the users in its cell), and a relay, in the case of multi-hop backhaul systems. 

The scaling of the per-connection (source-destination pair) throughput of wireless networks has been widely studied in the context of ad-hoc networks, which are conceptually closely related to the wireless backhaul system considered in this paper. While scalability is not achievable with single antenna transmission and multi-hop relaying \cite{GupKumJ2000,FraDouJ2007},  
there is definitely some hope in going to MIMO transmission. Besides, the current trend of moving towards higher and higher carrier frequencies (mm-wave communications)~\cite{MacZhaC2013,RapSunJ2013,AdhSafJ2013} makes it possible to pack more antennas in manageable form factors. For such systems, it is reasonable to envision a scenario where the number of antennas per BS, and hence the capacity of each backhaul link, increases with the 
network size. In this paper, we explore this networking paradigm in detail and show that it is possible to implement a scalable wireless backhaul network using MIMO transmission for each backhaul link, without violating any fundamental limits on the spatial DoFs.

As a disclaimer, we would like to point out that we are not envisaging a system where one should increase the number of antennas per BS with the addition of each new BS in the network in order to preserve scalability. Such a system would be highly impractical since it requires to change  the hardware at each BS as the network topology evolves. In contrast, as often happens in Information Theory, our analysis captures an ensemble of systems designed such that the number of antennas per BS is a function of the overall network size. By capturing this dependency, we can provide useful design guidelines and answers to several important questions, such as, how many antennas per BS are necessary for a network of given size in order to achieve a certain desired target backhaul rate per source-destination pair. 

\subsection{Related Work}  \label{related}

The study of the asymptotic throughput scaling laws of large wireless networks, where each node may be at the same time a source, a destination, and a relay node, has received a lot of attention  in the past decade (e.g., see~\cite{GupKumJ2000,FraDouJ2007,XieKumJ2004,OzgLevJ2007,FraMigJ2009,LeeChuJ2012,OzgLevJ2013}). While giving a precise account of all these results is out of the scope of this work, we summarize briefly the relevant ones. For concreteness, we use the following {\em ordering} notation in this section: given two functions $f$ and $g$, $f(n) = O\left(g(n)\right)$ if there exists a constant $c$ and integer $N$ such that  $f(n)\leq cg(n)$ for all $n>N$. Also, $f(n) = \Theta\left(g(n)\right)$ if $f(n) = O\left(g(n)\right)$ and~$g(n) = O\left(f(n)\right)$. We will revisit this notation in Section~\ref{sec:SysMod}.

Studies on capacity scaling laws of wireless networks have considered both {\em extended} and {\em dense} network models, where the former refers to the deployment of nodes
with constant spatial density, such that the network coverage area is $\Theta(n)$, with $n$ denoting the number of nodes, while the latter refers the case where the coverage area is $O(1)$ (constant with $n$), and the network density grows as $\Theta(n)$. In this paper we focus on a system where BSs are deployed with given (constant) density. Therefore, as far as the wireless backhaul is concerned, we are in the extended network regime. The seminal paper by Gupta and Kumar~\cite{GupKumJ2000}, successively 
refined by Franceschetti~{\em et al.}~\cite{FraDouJ2007},  showed that 
the sum capacity of such networks scales as $\Theta(\sqrt{n})$ for single-antenna wireless networks under 
a multi-hop decode and forward  relaying strategy.  
Since the number of source-destination pairs is $\Theta(n)$, this leads to  a per-connection throughput that decreases as $\Theta(1/\sqrt{n})$. 
Successively (see \cite{XieKumJ2004} and the summary of results in~\cite{OzgLevJ2007}), it was shown that
this behavior is information-theoretically order-optimal (over all possible relaying strategies) for power-law pathloss model 
with exponent $\alpha > 3$.  For pathloss exponents $2\leq\alpha<3$, \"{O}zg\"{u}r {\em et al.}~\cite{OzgLevJ2007} proposed a layered 
{\em hierarchical cooperation} scheme that can achieve sum throughput $\Theta(n^{1 - \epsilon})$ for arbitrarily small $\epsilon > 0$ 
and a sufficiently large number of hierarchical layers.
In contrast, by combining  Maxwell propagation laws with an information theoretic cut-set bound argument,  
Franceschetti~{\em et al.}~\cite{FraMigJ2009} showed that the total spatial DoFs of the network are limited by 
$\Theta(\sqrt{a}/\lambda)$,  where $\sqrt{a}$ here can be interpreted as network ``diameter''. Thus, for the extended model 
of area $a = \Theta(n)$, throughput $\Theta(\sqrt{n})$ is the best we can hope for.  This {\em dichotomy} of results was settled in the case of 
line-of-sight (LoS) propagation by Lee {\em et al.}~\cite{LeeChuJ2012} and by \"{O}zg\"{u}r {\em et al.}~\cite{OzgLevJ2013}, which showed 
that these results hold in different operational regimes. The key insight is provided by the achievable DoFs (equivalently, spatial multiplexing gain) 
of point-to-point MIMO channels with LoS propagation, characterized in \cite{LeeChuJ2012,OzgLevJ2013}. 
In the case of non-LoS {\em dense scattering} propagation, \"{O}zg\"{u}r {\em et al.}~\cite{OzgJohJ2010} 
have characterized different scaling regimes depending on the pathloss exponent and the typical nearest-neighbor SNR scaling exponent with $n$.  These regimes hold as long as $n < \sqrt{a}/\lambda$, i.e., when the network operates far from the electromagnetic 
propagation bottleneck of \cite{FraMigJ2009}.

In terms of the wireless backhaul network design, the focus of prior work has traditionally been on the cross layer optimization involving routing, scheduling and physical layer resource allocation. The main goal is to deliver data from the source BSs to a gateway node, which is connected to a wired backbone. A small sample of works in this direction is~\cite{GamSadC2004,CaoWanJ2007,AnaAraJ2008}. A more recent research direction has considered the performance of wireless backhaul links under practical considerations in the context of mm-wave communications. For instance,  Hur {\em et al.} \cite{HurKimJ2013} studied mm-wave beamforming for wireless backhaul links with emphasis on the effect of wind induced pole movement on beam alignment. On the contrary, we consider a purely wireless backhaul network in this paper, where none of the BSs have access to wired backhaul, and study its scalability as the number of BSs grow large. The main contributions are summarized next.

\subsection{Novelty and Main Outcomes} \label{novelty}

{\em Realistic setup for wireless backhaul networks.} There are at least three main characteristics of modern wireless backhaul networks that are captured in this work: (i) the size of urban small cell networks is quickly growing, (ii) there is a gradual shift towards higher transmission frequencies, and (iii) increasing 
maturity of mm-wave communications makes it possible to support large number of antennas at each BS, thus forming high capacity backhaul links amongst BSs. We model this networking paradigm as an {\em extended} network, where both the transmission frequency and the number of antennas per BS can scale with the network size. 

{\em Information theoretic bound.} In Section~\ref{sec:ITbounds}, we derive an information theoretic upper bound on the capacity of the wireless backhaul network. The idea is to use information cut-set bound over a cut that divides the network into two equal halves. The information flow across the cut is upper bounded by the capacity of the MIMO channel, where the BSs on the either side of the cut act as a distributed transmitter and receiver, respectively. The main technical arguments are based on a simple generalization of {\em geometric exponential stripping} technique of~\cite{FraJ2007} to channel matrices with complex-valued channel gains. An important consequence of this result is a lower bound on the number of antennas per BS required to ensure scalability.

{\em Short hop vs. long hop strategies.} In Sections~\ref{sec:LongHop} and~\ref{sec:ShortHop}, respectively, we consider two competing transmission strategies: (i) {\em long hop}: each source-destination pair minimizes the number of hops by sacrificing multiplexing gain and ideally achieving full power gain over each hop, and (ii) {\em short hop}: each source-destination pair communicates through a series of short hops, each achieving full multiplexing gain. While the long hop strategy may seem more reasonable, especially in the context of massive MIMO, where it is, in principle, possible to form thin beams in the direction of a far-off BS without creating excessive interference to its nearby BSs, we show that the short hop strategy is significantly efficient in terms of antenna requirement for throughput scalability. 
Hence, the relevance of works concerned with the {\em pointing problem} of narrow beams over long distances (long hops) in non-ideal conditions is, at best, questionable.

{\em System design insights.} The main design insight provided by our analysis is that it is possible to implement throughput-scalable wireless backhaul networks by forming high capacity MIMO links  between BSs, without violating any fundamental limits on the electromagnetic propagation bottleneck. Under our achievable strategy, the capacity of each short hop MIMO link should have DoFs that scale as the square-root of the number of BSs. Quantitative concrete examples showing the attractiveness of the advocated approach are provided in Remark \ref{rem:urban} and Section~\ref{sec:Discussion}. In Section \ref{sec:Discussion}, we also demonstrate how our analysis can be extended to more general networks with some of the BSs having access to wired backhaul.

\section{System Model} \label{sec:SysMod}

We consider a cellular network where the locations of the BSs are sampled from a homogeneous Poisson Point Process (PPP) $\Phi \subset \nbbR^2$ with density $\lambda_{\rm b}$ BSs per unit area. We further assume that none of the BSs have access to the conventional wired backhaul, and that all the data has to be communicated over wireless backhaul links. All the BSs share the same spectrum for their backhaul communication. As noted in the previous section, we do not consider the underlying cellular communication (uplink and downlink) between wireless users  and BSs. Assuming a well-planned and load balanced system, we can easily imagine that each  BS has to handle a fixed amount of traffic that does not grow with the overall number of BSs in the network. Our focus, here, is on the ability of a wireless backhaul network to relay such traffic.

For the scaling results, we consider the {\em random extended network model}, where we focus our attention on the box $B_n$ with size $\sqrt{n} \times \sqrt{n}$. The number of BSs lying in $B_n$ is a Poisson distributed random variable with mean $\lambda_{\rm b}n$. We are concerned with the asymptotic capacity scaling of the network formed by the BSs inside $B_n$ as $n \rightarrow \infty$.
In this regime, the box $B_n$ also grows, eventually encompassing all the points of $\Phi$, hence the name ``extended network''. Assuming uniform traffic, the source-destination pairs in $B_n$ are picked uniformly at random, such that each BS is a destination of exactly one source. 

For the wireless backhaul links, each BS has $\Psi(n)$ antennas, where $\Psi(\cdot): \nbbN^+ \rightarrow \nbbN^+$ is a monotonically non-decreasing function of $n$. Note that the antenna scaling assumption already appears in the literature, although in slightly different contexts, e.g., see~\cite{ShiJeoJ2011}. 
Furthermore, we assume that the physical dimensions (size) of antenna array does not change with $n$. 
Denoting the distance between the $k^{th}$ antenna of the transmitting BS to the $i^{th}$ antenna of the receiving BS by $d_{ik}$, 
the baseband channel gain $h_{ik}$ between these two antennas is
\begin{align}
h_{ik} = \sqrt{l(d_{ik})} \exp\left(j\theta_{ik}\right),
\label{eq:hik}
\end{align}
where $l(d_{ik}) = \min\{ 1,d_{ik}^{-\alpha} \}$ is a bounded power-law pathloss function with exponent $\alpha>2$, 
and $\theta_{ik}$ denotes phase rotation, which is typically a function of $d_{ik}$. Note that our analysis in Sections~\ref{sec:ITbounds} and~\ref{sec:LongHop} holds for any given $\{\theta_{ik}\}$, $1\leq i, k \leq \Psi(n)$, irrespective of their joint distribution and their dependence upon $\{d_{ik}\}$. 
We will, however, need to put more structure on $\{\theta_{ik}\}$ in Section~\ref{sec:ShortHop}, where we focus our attention on an achievability scheme under LoS propagation. In this case, we have $\theta_{ik}=\frac{ 2 \pi d_{ik}}{\lambda}$, where $\lambda$ is the transmission wavelength. Further details about the LoS model are intentionally delayed until Section~\ref{sec:ShortHop}, before which they are not needed. 

We denote the network throughput, i.e., total number of bits/sec successfully decoded at the destinations in $B_n$, by $T(n)$. The {\em worst-case} achievable rate per source-destination pair (over all such pairs) is denoted by $R(n)$. The following probabilistic version of the ordering notation is used~\cite{FraDouJ2007}. We write $f(n) = O(g(n))$ with high probability (w.h.p.) if $\exists$ a constant $K$ independent of $n$ such that $\lim_{n \rightarrow \infty} \nbbP(f(n) \leq K g(n)) = 1.$ Similarly, $f(n) = \Omega(g(n))$ if $g(n) = O(f(n))$. In the same spirit, any general event $\ncalA_n$ is said to occur w.h.p. if $\lim_{n\rightarrow \infty}\nbbP(\ncalA_n) = 1$. For notational simplicity, the bandwidth $W$ is assumed to be $1$ Hz and the noise power spectral density $N_0$ to be 1 watts/Hz. Also, each BS is assumed to have a maximum power constraint of $P$ watts. Since $R(n)$ is defined as the worst-case rate, it is easy to establish that $T(n) = \Omega(n R(n))$ and hence $R(n) = O\left( \frac{T(n)}{n} \right)$.

\section{Information Theoretic Upper Bound} \label{sec:ITbounds}

In this section, we derive an information-theoretic upper bound on the network throughput $T(n)$. 
We will make use of the following concentration Lemma.

\begin{lemma} \label{lem:Chernoff_Poisson}
For a homogeneous PPP $\Phi \subset \nbbR^d$ with density $\lambda_{\rm b}$, let $N(\ncalA)$ be the number of points in any measurable set $\ncalA \subset \nbbR^d$. Denote the Lebesgue measure of $\ncalA$ by $|\ncalA|$. We have
\begin{align}
&\lim_{|\ncalA| \rightarrow \infty} \nbbP(N(\ncalA) \geq 2\lambda_{\rm b} |\ncalA|) = 0, \label{eq:Chernoff_Poisson_UB}\\
&\lim_{|\ncalA| \rightarrow \infty} \nbbP\left(N(\ncalA) \leq \frac{\lambda_{\rm b}}{2} |\ncalA|\right) = 0. \label{eq:Chernoff_Poisson_LB}
\end{align}
\end{lemma}

\begin{IEEEproof}
See Appendix~\ref{app:Chernoff_Poisson}.
\end{IEEEproof}

\begin{figure}
\centering
\includegraphics[width=\columnwidth]{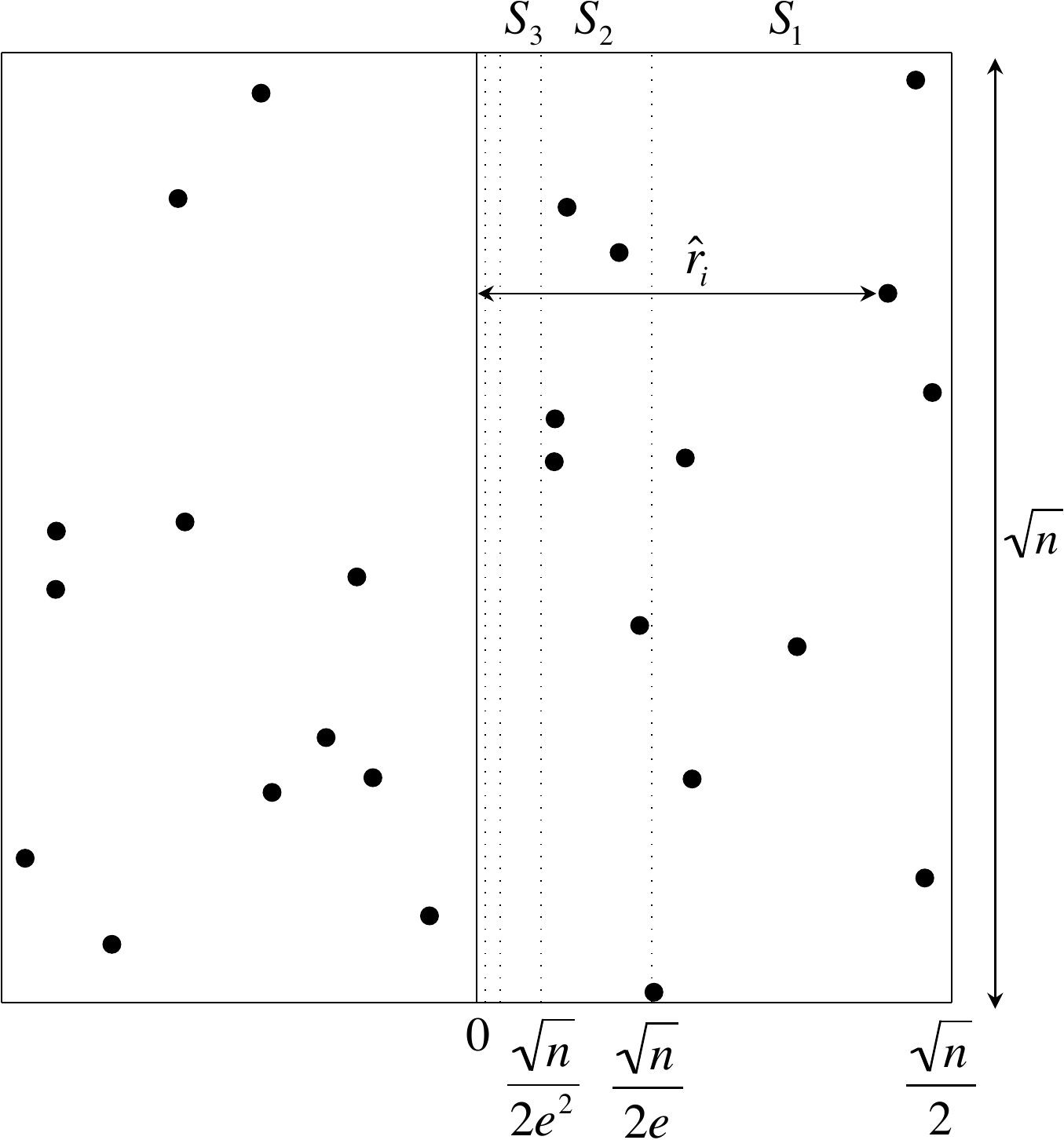}
\caption{The setup to derive information-theoretic upper bounds. The vertical strips are denoted by $S_i$, where $1\leq i \leq \lfloor \log \frac{\sqrt{n}}{2} \rfloor + 1$.}
\label{fig:UBproof_fig}
\end{figure}

The upper bound is obtained through a cut-set bound argument. We partition the box $B_n$ into two halves, each with side lengths $\sqrt{n} \times \sqrt{n}/2$, as shown in Fig.~\ref{fig:UBproof_fig}. We will study the information flow across the common edge of the two boxes, i.e., this edge acts as a cut. By Lemma~\ref{lem:Chernoff_Poisson}, we have that w.h.p. there are less than $\lambda_{\rm b}n$ points in each half-box. Since there are $O(n)$ source-destination pairs that need to transmit across this cut, the upper bound on the information flow across the cut also gives an upper bound (in order) for $T(n)$, from which the upper bound on source-destination rate $R(n)$ directly follows. Note that the information flow across the cut is upper bounded by the capacity of the effective MIMO channel, say $C_n$, with the BSs to the left of the cut operating as an effective transmitter and the BSs to the right as a receiver. Since we are interested in an upper bound, we can assume that there are exactly $\lambda_{\rm b}n$ BSs on either side of the cut. Further, since the value of $\lambda_{\rm b}$ does not impact scaling results as long as it is finite, we fix it to $1$ in this section 
for ease of notation. The $n\Psi(n) \times n\Psi(n)$ effective channel matrix is given by 
\begin{align}
\nbH_{\rm eff} = \left[
\begin{array}{cccc}
\nbH_{11} & \nbH_{12} & \ldots & \nbH_{1n}\\
\nbH_{21} & \nbH_{22} & \ldots & \nbH_{2n}\\
\vdots       & \vdots        & \ddots & \vdots\\
\nbH_{n1} & \nbH_{n2} & \ldots & \nbH_{nn}
\end{array}
\right],
\label{eq:ChanMatGen}
\end{align}
where $\nbH_{ik}$ is a $\Psi(n) \times \Psi(n)$ channel matrix from $k^{th}$ BS from the left of the cut to the $i^{th}$ BS to the right of the cut. 
Now, denoting the transmit symbol covariance matrix by $\nbQ$, we can write
\begin{align}
C_n &= \max_{\substack{\nbQ \succ 0 \\ \Tr(\nbQ)\leq nP}} \log \det \left( \nbI + \nbH_{\rm eff} \nbQ \nbH_{\rm eff}^\dag \right) \nonumber \\
&\stackrel{(a)}{\leq}  \log \det \left( \nbI + nP\nbH_{\rm eff} \nbH_{\rm eff}^\dag \right) \nonumber \\
&\stackrel{(b)}{\leq}  \sum_{i=1}^{n\Psi(n)} \log \left( 1 + nP(\nbH_{\rm eff} \nbH_{\rm eff}^\dag)_{ii} \right)
\label{eq:Cn_UB}
\end{align}
where $(a)$ follows by the fact that $\log\det(\cdot)$ is monotonically increasing on the cone of positive semidefinite Hermitian matrices, and
for any $\nbQ$ satisfying $\Tr(\nbQ)\leq nP$ we have $\nbQ \leq nP \nbI$,\footnote{For matrices $\nbA$ and $\nbB$ we write
$\nbA \leq \nbB$ to indicate that the difference $\nbB - \nbA$ is positive semi-definite.}
and $(b)$ follows from Hadamard inequality. For the notational ease, we assume that the distances between any pair of transmit and receive antennas between $k^{th}$ BS on the left 
(transmitter) and the $i^{th}$ BS on the right (receiver) are the same and equal to $r_{ik}$. 
The main idea now is to derive tight bounds on $(\nbH_{\rm eff} \nbH_{\rm eff}^\dag)_{ii}$ using the geometric properties of the point process that determine the distances of BSs from the cut. The bound is based on the tools developed in~\cite[Theorem 5.4.4]{FraMeeB2007}, \cite{FraJ2007}, where a similar bound is derived for a single antenna network. A key difference is that the ``mirroring argument'' used in~\cite{FraJ2007,LevTelJ2005} to establish equivalence between singular and eigenvalues of the channel matrix is not directly applicable in our case due to complex-valued channel gains. 
However, as discussed in the sequel, especially Appendix~\ref{app:Cs_UB}, this requires only a few technical adjustments of the proof in~\cite{FraJ2007}.

We order the BSs on both sides of the cut by their respective distances from the cut. The distance of BS $i$ from the cut is denoted by 
$\hat{r}_i$. To get a tight bound on $(\nbH_{\rm eff} \nbH_{\rm eff}^\dag)_{ii}$, we use the {\em exponential stripping} technique introduced in~\cite{FraJ2007}. 
As shown in Fig.~\ref{fig:UBproof_fig}, both the half-boxes on the either side of the cut are partitioned into $\lfloor \log \frac{\sqrt{n}}{2} \rfloor + 1$ vertical strips $S_i$. For $1\leq i \leq \lfloor \log \frac{\sqrt{n}}{2} \rfloor$, the minimum distance of the BSs lying in $S_i$ from the cut is $\frac{\sqrt{n}}{2e^i}$, which will be used to upper bound $(\nbH_{\rm eff} \nbH_{\rm eff}^\dag)_{ii}$. For $i = \lfloor \log \frac{\sqrt{n}}{2} \rfloor + 1$, i.e., the vertical strip closest to the cut, we will simply upper bound the path-loss by $1$. Denoting the number of BSs in $S_i$ by $X(S_i)$, the following holds w.h.p. $\forall\ i$
\begin{align}
X(S_i) \leq \frac{n} {e^i} \left( e - 1\right).
\label{eq:XSi_LB}
\end{align} 
The result for $1\leq i \leq \lfloor \log \frac{\sqrt{n}}{2} \rfloor$ follows directly from Lemma~\ref{lem:Chernoff_Poisson} and for $i = \lfloor   \log \frac{\sqrt{n}}{2} \rfloor + 1$ we have 
\begin{align}
&\lim_{|\ncalA| \rightarrow \infty} \nbbP(N(\ncalA) > 2(e-1) \lambda |\ncalA|)  \nonumber \\
&\leq \lim_{|\ncalA| \rightarrow \infty} \nbbP(N(\ncalA) > 2\lambda |\ncalA|) \stackrel{(a)}{=} 0, 
\end{align}
where $(a)$ follows again from Lemma~\ref{lem:Chernoff_Poisson}. From \eqref{eq:Cn_UB}, we get
\begin{align}
C_n &\leq \Psi(n) \sum_{i=1}^{\log \frac{\sqrt{n}}{2} + 1} X(S_i) \log \left( 1 + nP(\nbH_{\rm eff} \nbH_{\rm eff}^\dag)_{ii} \right) \\
&\leq \left( e - 1\right) \Psi(n) \sum_{i=1}^{ \log \frac{\sqrt{n}}{2} + 1} \frac{n} {e^i}  \log \left( 1 + nP(\nbH_{\rm eff} \nbH_{\rm eff}^\dag)_{ii} \right),
\label{eq:Cn_powlawpl_1}
\end{align}
where we expressed $\lfloor  \log \frac{\sqrt{n}}{2} \rfloor$ simply as $\log \frac{\sqrt{n}}{2}$ without  compromising our results 
because we are interested in the behavior for $n \rightarrow \infty$. Recall that index $i$ in $(\nbH_{\rm eff} \nbH_{\rm eff}^\dag)_{ii}$ 
corresponds to all the BSs in the vertical strip $S_i$, which have the common upper bound
\begin{align}
(\nbH_{\rm eff} \nbH_{\rm eff}^\dag)_{ii} &= \Psi(n) \sum_{k=1}^n l(r_{ik}) \nonumber\\
&\stackrel{(a)}{\leq} \Psi(n) \sum_{k=1}^n l(\hat{r}_i) = n \Psi(n) l(\hat{r}_i) \nonumber\\
&\stackrel{(b)}{\leq} \left\{ \begin{array}{cc}
n \Psi(n) \hat{r}_i^{-\alpha} & 1\leq i \leq  \log \frac{\sqrt{n}}{2} \\
n\Psi(n) & i = \log \frac{\sqrt{n}}{2} + 1
\end{array}
\right.  \nonumber\\
&\stackrel{(c)}{\leq} \left\{ \begin{array}{cc}
n^{1-\frac{\alpha}{2}} \Psi(n)2^{\alpha} e^{i\alpha} & 1\leq i \leq  \log \frac{\sqrt{n}}{2} \\
n\Psi(n) & i = \log \frac{\sqrt{n}}{2} + 1
\end{array}
\right.,
\label{eq:HH_ii_tight}
\end{align}
where $(a)$ follows from the fact that $\hat{r}_i \leq r_{ik}$ and $l(\cdot)$ is a non-increasing function, $(b)$ from the fact that 
for $1\leq i \leq \log \frac{\sqrt{n}}{2}$, $\hat{r}_i$, which is the distance of the BS lying in $S_i$ from the cut, 
is lower bounded by $\frac{\sqrt{n}}{2e^i} \geq 1$, which implies $l(\hat{r}_i) = \hat{r}_i^{-\alpha}$, and for $i = \log \frac{\sqrt{n}}{2}+1$, we simply upper bound $l(\hat{r}_i)$ by $1$, and $(c)$ follows by lower bounding $\hat{r}_i$. Substituting \eqref{eq:HH_ii_tight} in \eqref{eq:Cn_powlawpl_1}
\begin{align}
C_n &\leq \left( e - 1\right) \Psi(n) \sum_{i=1}^{ \log \frac{\sqrt{n}}{2}} \frac{n} {e^i}  \log \left( 1 + P n^{2-\frac{\alpha}{2}} \Psi(n)2^{\alpha} e^{i\alpha} \right)  \nonumber \\
&+ 2 \frac{ e - 1}{e} \Psi(n)  \sqrt{n}  \log \left( 1 + P n^2 \Psi(n) \right),
\label{eq:Cn_powlawpl_2}
\end{align}
where the last term is $O(\Psi(n)\sqrt{n}\log(n^2 \Psi(n)))$. Denoting the summation in the first equation by $C_{\rm s}$, 
we can express~\eqref{eq:Cn_powlawpl_2} as
\begin{align}
C_n &\leq \left( e - 1\right) \Psi(n) C_{\rm s} + O(\Psi(n)\sqrt{n}\log(n^2 \Psi(n))).
\label{eq:Cn_powlawpl_3}
\end{align}
After some effort, we can prove the following result on the scaling of $C_{\rm s}$.
\begin{lemma} \label{lem:Cs_UB}
For path-loss exponent $\alpha > 2 \left(2 + \log_n \Psi(n) \right)$
\begin{align}
C_{\rm s} &= O\left(\sqrt{n} n^{\frac{2}{\alpha}} \Psi(n)^{\frac{1}{\alpha}} \log n\right).
\end{align}
\end{lemma}

\begin{IEEEproof}
See Appendix~\ref{app:Cs_UB}.
\end{IEEEproof}

From Lemma \ref{lem:Cs_UB}, it is clear that the first term of~\eqref{eq:Cn_powlawpl_3} scales as $O\left(\sqrt{n} n^{\frac{2}{\alpha}} \Psi(n)^{1+\frac{1}{\alpha}} \log n\right)$, which represents the dominating term. This leads to the following upper bound on $T(n)$.
\begin{theorem} \label{thm:IT_UB_main}
For path-loss exponent $\alpha > 2 \left(2 + \log_n \Psi(n) \right)$
\begin{align}
T(n) &= O\left(\sqrt{n} n^{\frac{2}{\alpha}} \Psi(n)^{1+\frac{1}{\alpha}} \log n\right).
\end{align}
\end{theorem}


\begin{cor}
The number of antennas needed per-BS in order to achieve $R(n) = O(1)$ for $\alpha > 4$ is
\begin{align}
\Psi(n) = \Omega \left ( \left[n^{\frac{1}{2} - \frac{2}{\alpha}}(\log n)^{-1} \right]^{\frac{\alpha}{1+\alpha}} \right ).
\end{align}
\end{cor}

\begin{remark}[Scalability]
The above Corollary should be interpreted as a lower bound on the number of antennas per BS needed for scalability. 
In particular, for high attenuation regime, we need to scale antennas almost as $\sqrt{n}$ to make the backhaul network scalable. 
In fact, in Section~\ref{sec:ShortHop}, $\Psi(n)=\sqrt{n}$ is shown to achieve scalability for any $\alpha>2$ in LoS MIMO networks, 
thus showing that there exists a regime of BS physical sizes, inter-BS distances, and high frequency, where a scalable wireless backhaul 
can be effectively implemented with short hops, each achieving high MIMO multiplexing gain.
\end{remark}

\section{Long Hops: Beamforming} \label{sec:LongHop}

In this section, we consider a transmission strategy where each BS uses all its antennas for beamforming, i.e., 
it transmits a single data stream to the farthest possible BS in the direction of its destination in order to minimize the number of hops. 
Our goal is to find an upper bound on the achievable $R(n)$ under this strategy, as a function of $\Psi(n)$ and $\alpha$. 
Recall that for a given link with  transmitter-receiver separation of $d > 1$, and $\Psi(n)\times \Psi(n)$ channel matrix $\nbH$ with 
entries given by~\eqref{eq:hik}, the maximum rate achievable for a single stream under eigen-beamforming is
\begin{align}
{\rm Rate}(d) &\stackrel{(a)}{=} \log \left(1+ P \lambda_{\max}(\nbH \nbH^{\dag}) \right) \nonumber \\
&\stackrel{(b)}{\leq}  \log \left(1+ P d^{-\alpha} \Psi(n)^2 \right),
\label{eq:rate_BF}
\end{align}
where $\lambda_{\max}(\nbH \nbH^{\dag})$ in $(a)$ is the maximum eigenvalue of $\nbH \nbH^{\dag}$, 
and $(b)$ follows from the fact that $\lambda_{\max}(\nbH \nbH^{\dag}) \leq  \Tr(\nbH \nbH^{\dag}) = d^{-\alpha} \Psi(n)^2$. 
To minimize the number of hops for a given source-destination pair, our goal is to maximize distance $d$ for each hop keeping the transmission rate \eqref{eq:rate_BF} constant. 
Assuming a minimum target received power equal to some value $P_0$, we have
\begin{align}
P d^{-\alpha} \Psi(n)^2 \geq P_0 \Rightarrow d \leq  \left( \frac{P}{P_0} \right)^{\frac{1}{\alpha}} \Psi(n)^\frac{2}{\alpha} = d_{\rm c}.
\end{align}
Given $d_{\rm c}$, we need a lower bound on the source-destination separation in order to lower bound the number of hops needed. 
This is provided by the following Lemma.

\begin{lemma}[Lower bound on source-destination separation] \label{lem:LB_D}
The source-destination separation of a randomly chosen pair in $B_n$ is $\Omega(n^{\frac{1}{2} - \epsilon})$ w.h.p., where $\epsilon>0$.
\end{lemma}

\begin{IEEEproof}
See Appendix~\ref{app:LB_D}.
\end{IEEEproof}

Then, the number of hops $N_{\rm h}$ required for a randomly chosen source-destination pair  is lower bounded as
$N_{\rm h} = \Omega \left(\frac{n^{\frac{1}{2} - \epsilon}}{d_{\rm c}} \right).$
From Lemma~\ref{lem:Chernoff_Poisson}, we have that the number of BSs in $B_n$ is $\Omega(n)$ and therefore the number of source-destination pairs is also $\Omega(n)$. As a result, a lower bound on the total number of hops needed in the network is 
$n N_{\rm h} = \Omega \left(\frac{n^{\frac{3}{2} - \epsilon}}{d_{\rm c}} \right).$ 
Recall that each hop achieves a fixed target receiver power $P_0$, therefore it can support a fixed peak rate, which needs to be shared among all the connections handled by the ``bottleneck'' BS.
Hence, there is at least one BS that has to relay $\Omega \left(\frac{n^{\frac{1}{2} - \epsilon}}{d_{\rm c}} \right)$ connections. Therefore, the rate per source-destination pair is upper bounded by $R(n) = O\left(d_{\rm c} n^{-\frac{1}{2} + \epsilon} \right)$, which leads to the following main result.

\begin{theorem}[Long hop] \label{thm:long-hop}
For the long hop strategy discussed in this section
\begin{align}
R(n) = O\left(\frac{\Psi(n)^{\frac{2}{\alpha}}} {n^{\frac{1}{2} - \epsilon}} \right).
\end{align}
\end{theorem}

\begin{remark}[Scalability under long-hop strategy] \label{rem:long-hop}
From Theorem~\ref{thm:long-hop}, it is clear that in order to achieve $R(n) = O(1)$, we need $\Psi(n) = \Omega(n^{\frac{\alpha}{4} - \epsilon})$, which for a vanishingly small $\epsilon$ and $\alpha>2$ is always higher than $\sqrt{n}$. In fact, for $\alpha=4$, we need to scale the number of 
antennas almost linearly with $n$.
\end{remark}

\section{Short Hops in LoS: Spatial Multiplexing}  \label{sec:ShortHop}

\begin{figure}
\centering
\includegraphics[width=\columnwidth]{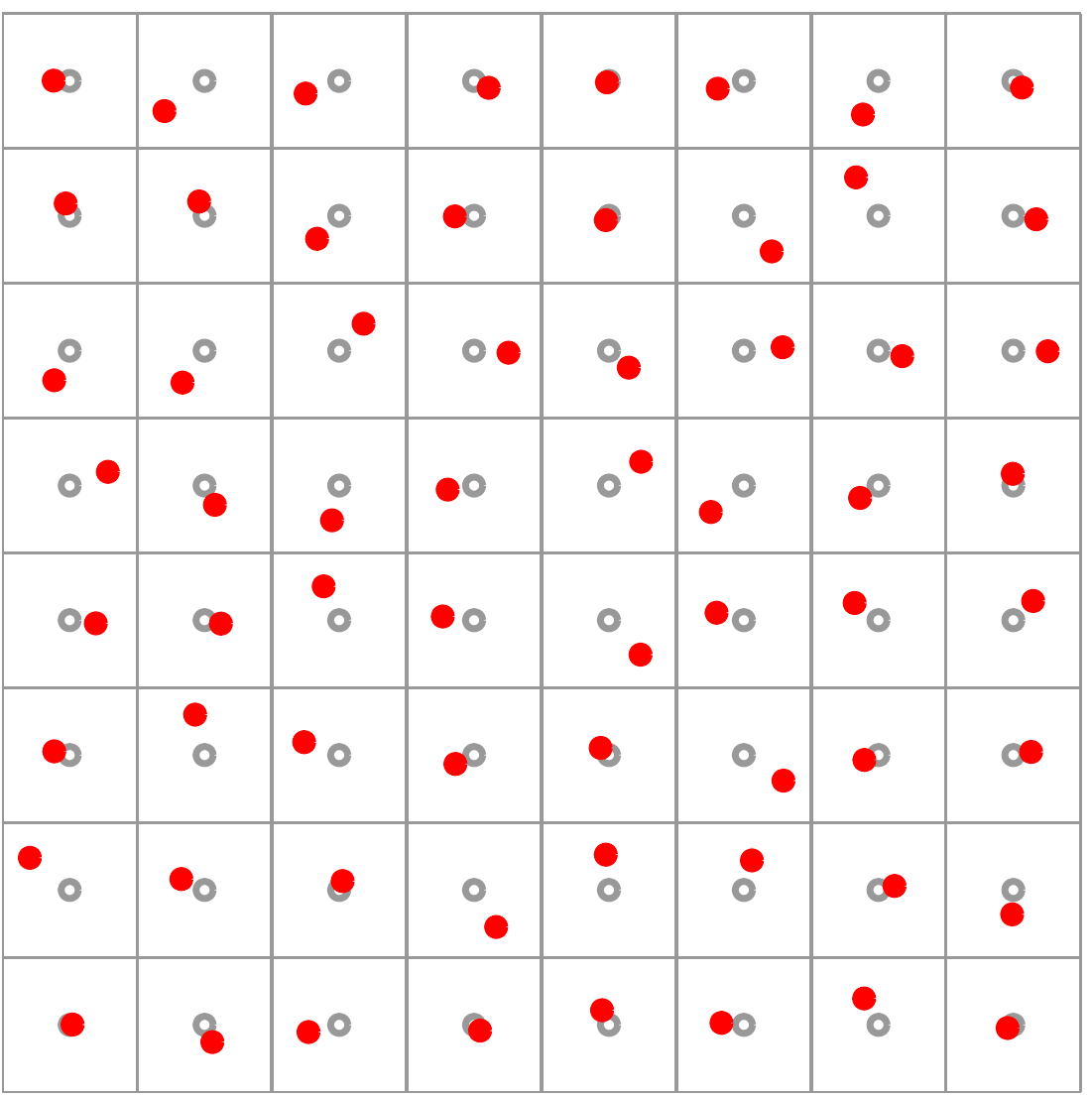}
\caption{Perturbed lattice model for the BS locations. The lattice and the BS locations are denoted by hollow and filled circles, respectively.}
\label{fig:GridLocations}
\end{figure}

In this section, we consider the other extreme where the data for each source-destination pair is communicated through a series of {\em short hops}, also termed as {\em information highways} in the context of random networks~\cite{FraDouJ2007}.
For the ease of exposition, we introduce the key ideas using a simpler, but still meaningful, model shown in Fig.~\ref{fig:GridLocations}, 
where the BS locations are given by a perturbed square lattice. The distance between the closest lattice points is assumed to be a constant 
$c = \lambda_{\rm b}^{-\frac{1}{2}}$ in order to preserve the BS density $\lambda_{\rm b}$ as in  Section~\ref{sec:SysMod}. 
After establishing the results for this simpler model, we will generalize them to the random network using percolation theory 
arguments of~\cite{FraDouJ2007}. By choosing this simpler model for exposition, we avoid repeating many key arguments of~\cite{FraDouJ2007}, e.g., the construction of bond percolation model,  while still conveying the main message.

By short hop, we specifically refer to the communication link between two neighboring BSs, each lying in adjacent 
small squares in Fig.~\ref{fig:GridLocations}.  As reviewed in Section \ref{related}, in classical single-antenna wireless networks
the short-hop  strategy achieves $R(n) = \Omega\left(\frac{1}{\sqrt{n}}\right)$. 
The main intuition behind this result can be explained in terms of {\em horizontal} and {\em vertical} routes, 
formed by rows and columns of small squares in Fig.~\ref{fig:GridLocations}, running from left to right or from top to bottom edges of $B_n$. 
Note that under single-antenna transmission, both the horizontal and vertical routes achieve rate $\Omega(1)$ if the interference power at each BS is bounded. The result now follows from the fact that any destination can be reached from its source by following a simple routing strategy, where data is first sent over a horizontal route until it reaches the column where its destination lies, after which it is sent over that vertical route. In other words, each route is shared by $\sqrt{n}$ connections, which implies that the rate per connection is $R(n) = \Omega\left(\frac{1}{\sqrt{n}}\right)$. A natural question to ask now is what happens when these short hops are MIMO links capable of transmitting multiple independent streams by spatial multiplexing. If the number of streams remains constant independent of $n$, each route achieves a higher but constant rate, which does not affect the scaling results. However, if they scale up as $\Omega(\sqrt{n})$, the rate of each route scales as $\sqrt{n}$, which implies a per source-destination rate of $R(n) = \Omega(1)$. Therefore, the main goal of this section is to find $\Psi(n)$ which enables each short hop MIMO link to achieve rate, say $R_{\rm sh} (n)$, of $\Omega(\sqrt{n})$ in the presence of interference due to other simultaneous MIMO transmissions.

Since this section deals with an achievability result, we need to be careful with the channel model in order to avoid conflicting conclusions. We explicitly consider LoS propagation due to the following reasons: 
(i) LoS propagation model is consistent with our general channel gain definition given by~\eqref{eq:hik};
(ii) it is relevant in the context of high carrier frequencies (mm-waves range) for which ``rich scattering'' has not been observed in channel measurements. In fact, at such high frequencies, the rank of non-LoS channel matrices
is limited by the number of dominant scatterers rather than by the number of antennas (see the recent work in \cite{AdhSafJ2013} and references therein);
(iii) LoS propagation for the backhaul can be obtained by roof-top mounted antenna arrays, especially in sub-urban areas where homes have roughly the same height; (iv) the structure of LoS channel matrices yield a natural collapse of channel DoFs as a function of geometry, thereby avoiding the annoying 
dichotomy of results that occurs under the rich scattering assumption (discussed in Section~\ref{related}).
Before going into more technical details, we  review the key results for LoS propagation in the context of our model.

\subsection{LoS Propagation} \label{subsec:LoSPropagation}

We assume that the antennas at each BS node are uniformly distributed in squares of side $\sqrt{a}$, 
significantly smaller than the inter-BS distance. In practice, for  finite number of antennas, the orientation 
of arrays and the placement of antenna elements can be carefully optimized to maximize the rate achievable by a LoS MIMO link, 
e.g., see~\cite{BohOrtJ2007}. For $\Psi(n) \rightarrow \infty$, the achievable spatial DoFs for a LoS $\Psi(n) \times \Psi(n)$ MIMO channel 
were independently derived in~\cite{LeeChuJ2012,OzgLevJ2013} to be
\begin{align}
\left\{  \begin{array}{cc}
\min\left\{\Psi(n), \frac{\sqrt{a}}{\lambda}  \right\}, & {\rm when}\ d \in \left[1, \sqrt{a} \right]\\
\min\left\{\Psi(n), \frac{a}{\lambda d}  \right\}, & {\rm when}\ d \in \left(\sqrt{a}, \frac{a}{\lambda} \right]\\
1, & {\rm when}\ d \in \left(\frac{a}{\lambda}, \infty \right),
\end{array}
\right.
\label{eq:DoF}
\end{align}
where  $\lambda$ denotes the carrier wavelength and $d$ is the transmitter-receiver distance. 
A matching upper bound (within logarithmic factors) on the DoFs has been  recently derived in~\cite{DesLevC2013} under an 
element-by-element approximation of the LoS MIMO channel matrix by another random matrix. 
As discussed in the sequel, a particular case of interest for short hop strategy is when a LoS MIMO link can achieve DoFs $=\Psi(n)$ for all 
$d \leq d_{\max}$, for some constant $d_{\max}$ independent of $n$. From \eqref{eq:DoF}, it is clear that for $d_{\max} \leq \sqrt{a}$ 
this can be achieved by: (i) fixing $\lambda$ and scaling $\sqrt{a} \propto \Psi(n)$, or (ii) fixing $a$ and scaling $\lambda \propto \Psi(n)^{-1}$. In a realistic urban backhaul network, $d_{\max}$ would correspond to the separation between closest BSs and would be of the order of 100s of meters. Therefore, for reasonable form factors of the BSs, $d_{\max} \leq \sqrt{a}$ will never hold in practice. Now, focusing on the more realistic case of $d_{\max} > \sqrt{a}$, DoFs $=\Psi(n)$ can be achieved by: (i) fixing $\lambda$ and scaling $a \propto \Psi(n)$, or (ii) fixing $a$ and scaling $\lambda \propto \Psi(n)^{-1}$. While it is not possible to keep growing the size of BS with $\Psi(n)$, it is indeed possible to assume $\lambda \propto \Psi(n)^{-1}$, which is consistent with the ongoing migration towards higher transmission frequencies~\cite{RapSunJ2013}. An example of a possible practical layout is given next.

\begin{remark}[Sub-urban small-cell network] \label{rem:urban}
Consider an urban wireless backhaul network with $d_{\max} = 100$m. Also consider $\Psi(n) = 64$, which we know is possible with the 
current technology~\cite{BleJ2013}. To achieve full DoFs at a carrier frequency of $30$GHz, i.e., $\lambda = 1$cm, we need $a = \Psi(n)\lambda d_{\max} = 64$m$^2$, which means a square array of side length $8$m. This yields a quite practical setup, where
$\Psi(n)$ radiating elements for each BS can be integrated into roof or other architectural elements in order to enable a high-throughput 
wireless backhaul connectivity. Notice also that these antennas may be for backhaul-only purposes, since the cellular uplink and 
downlink can be handled at different frequencies through a different geometry (involving both BSs and users). 
\end{remark}

The assumption $\lambda \propto \Psi(n)^{-1}$ requires a careful treatment of link budget and pathloss, which we do next. 
In LoS conditions, the channel gain between two single-antenna nodes with separation $d\gg \lambda$ is~\cite{LeeChuJ2012}
\begin{align}
h &= \sqrt{G_{\rm T} G_{\rm R}} \frac{\lambda}{4\pi d} \exp\left(-j \frac{2\pi}{\lambda} d \right) \nonumber \\
&\stackrel{(a)}{=}  \frac{\sqrt{G_{\rm T} G_{\rm R}}}{4\pi d \Psi(n)} \exp\left(-j 2\pi d \Psi(n) \right),
\label{eq:h_single}
\end{align}
where $G_{\rm T}$ and $G_{\rm R}$ are the antenna gains at the transmitter and the receiver, respectively, and $(a)$ follows from $\lambda \propto \Psi(n)^{-1}$ with proportionality constant of 1 (choice of constant does not matter for the sake of scaling laws). 
In the context of our model, $G_{\rm T}$ and $G_{\rm R}$ refer to the gains of a single radiating element, not to the array gain of the $\Psi(n)$-element array.
As discussed in~\cite[Section 4.10]{ProSalB2008}, for a radiating element of physical area $A$, the antenna gain is
$G \propto \frac{A}{\lambda^2}$. Scaling up the number of array elements as $\Psi(n)$, while keeping the total size of the array constant, 
implies $A \propto \Psi(n)^{-1}$. This along with the fact that $\lambda \propto \Psi(n)^{-1}$ implies that both
$G_{\rm T}$ and $G_{\rm R}$ increase linearly with $\Psi(n)$. Substituting this back in \eqref{eq:h_single} for both $G_{\rm T}$ and $G_{\rm R}$ 
(within an irrelevant proportionality constant which is taken equal to 1 for simplicity of notation), we get
\begin{align}
h =  (4 \pi d)^{-1} \exp\left(-j 2\pi d \Psi(n)\right).
\label{eq:h_eq_los}
\end{align}
Note that, although uncommon, the scaling of frequency with $n$ has been considered before~\cite{LeeChuJ2012}. 
In this section, we adopt a slightly modified version of ~\eqref{eq:h_eq_los}. In particular, the gain between the $k^{th}$ antenna of the 
transmitter and the $i^{th}$ antenna of the receiver is
\begin{align}
h_{ik} =  \min\{d_{ik}^{-\frac{\alpha}{2}},1\} \exp\left(-j 2\pi d_{ik} \Psi(n)\right),
\label{eq:h}
\end{align}
where the pathloss exponent $\alpha > 2$ captures larger attenuation that may result from roof-top diffraction~\cite{OzgLevJ2013}, 
and $\min\{d_{ik}^{-\frac{\alpha}{2}},1\}$ avoids singularity at the origin. The constant $\frac{1}{4\pi}$ is irrelevant for the scaling results 
and it is ignored in~\eqref{eq:h}. For notational simplicity, we assume amplitude term $\min\{d_{ik}^{-\frac{\alpha}{2}},1\}$ in~\eqref{eq:h} to be the same and equal to $\min\{d^{-\frac{\alpha}{2}},1\}$ for all the transmit-receive antenna pairs for a given link due to the assumption that the 
BS linear size $\sqrt{a}$  is significantly  smaller than the separation between two BSs (see Remark \ref{rem:urban}). 
The LoS model~\eqref{eq:h} is consistent with the general model given by~\eqref{eq:hik}.

\subsection{Throughput for the Perturbed Lattice Model} \label{sec:ShortHop_Lattice}

The main challenge in analyzing $R_{\rm sh} (n)$ is the presence of LoS interference originating from other simultaneous MIMO transmissions. 
A closely related problem has recently been studied in the literature as a part of the {\em hierarchical cooperation strategy} for ad hoc networks, where one of the intermediate steps is to derive the rate achievable by a distributed MIMO transmission in the presence of interference from other simultaneous 
MIMO transmissions~\cite{OzgLevJ2007}. While the procedure to handle this interference under i.i.d. MIMO channels is well understood, 
it is not the case when the interfering MIMO links are LoS~\cite{LeeChuJ2012,OzgLevJ2013}. This problem is rigorously treated in~\cite{LeeChuJ2012}, where the distributed MIMO link rate is derived by explicitly considering LoS MIMO interfering links, see~\cite[Lemma 2]{LeeChuJ2012}. 
In this section, we take an alternate route and study the scaling of $\nbbE[R_{\rm sh} (n)]$, where expectation is over the (random) antenna locations 
at each BS. We show that $\Psi(n) = \sqrt{n}$ is sufficient to achieve $\nbbE[R_{\rm sh} (n)] = \Omega(\sqrt{n})$. 
Our analysis is considerably simpler and involves a direct bound on the interference power, reducing the problem to finding the spatial 
DoFs of a single LoS MIMO link in isolation, which is given by~\eqref{eq:DoF}. We then remark on the connections of this result with 
the scaling of $R_{\rm sh}(n)$.  We begin with the following Lemma.

\begin{lemma} \label{lem:sh_uniform}
For any continuous random variable $D\geq 0$
\begin{align}
\lim_{\Psi(n) \rightarrow \infty} \nbbE \left[ \exp\left(-j 2\pi D \Psi(n) \right) \right] = 0.
\end{align}
\end{lemma}

\begin{IEEEproof}
See Appendix~\ref{app:sh_uniform}.
\end{IEEEproof}

Assume $c\sqrt{5} \leq d_{\max}$ so that each short-hop achieves full spatial DoFs in the absence of interference (recall that $c$ is the 
distance between adjacent points in the lattice of Fig.~\ref{fig:GridLocations}). 
Denote the LoS MIMO channel matrix of the desired link by $\nbH$ and that of the $i^{th}$ interferer to the desired receiver 
by $\nbH^{(i)}$. Similarly, denote the transmit symbol of the desired and $i^{th}$ interfering transmitters by $\nbx$ and $\nbx^{(i)}$, respectively. 
We further assume that each transmitter distributes equal power across antennas, i.e., $\nbbE[\nbx \nbx^{\dag}] = \nbbE[\nbx^{(i)} \nbx^{(i)\dag}] = \frac{P}{\Psi(n)} \nbI$. For worst case analysis, we assume all the nodes are transmitting.\footnote{We could have used a reuse factor larger than 1 
to control interference temperature, e.g., see~\cite{OzgLevJ2007}, but this does not affect our scaling results because of the bounded pathloss model.} 
Denoting the set of interferers by $\ncalI$, and the noise vector by $\nbz$, the received signal at the desired receiver is
\begin{align}
\nby = \nbH \nbx + \sum_{i \in \ncalI} \nbH^{(i)} \nbx^{(i)} + \nbz.
\end{align}
The rate achievable by a short-hop link can be lower-bounded as
\begin{align}
R_{\rm sh} (n) \geq \log \det \left( \nbI + \frac{P}{\Psi(n)} \nbR^{-1} \nbH \nbH^{\dag} \right),
\label{eq:Rshn}
\end{align}
where $\nbR$ is the covariance matrix of noise-plus-interference observed at the desired 
receiver, given by
\begin{align}
\nbR = \nbI + \frac{P}{\Psi(n)}\sum_{i \in \ncalI}  \nbH^{(i)} \nbH^{(i)\dag}.
\label{eq:R_main}
\end{align}
Since the antenna locations of the interfering transmitters only affect \eqref{eq:Rshn} through $\nbR$, by applying iterated expectation and 
Jensen's inequality we have
\begin{align}
\nbbE[R_{\rm sh} (n)] \geq \nbbE \left [ \log \det \left( I + \frac{P}{\Psi(n)} \nbbE[\nbR]^{-1} \nbH \nbH^{\dag} \right) \right ],
\label{eq:RshnExp}
\end{align}
where, with a slight abuse of notation, the inner expectation is with respect to the antenna locations of the interfering transmitters 
conditioned on the antenna locations of the receiver, and the outer expectation is with respect to the antenna locations of the receiver and the 
intended transmitter. Now the goal is to upper bound the inner (conditional) expectation $\nbbE[\nbR]$, which is
\begin{align}
\nbbE[\nbR] = \nbI + \frac{P}{\Psi(n)}\sum_{i \in \ncalI}  \nbbE \left[ \nbH^{(i)} \nbH^{(i)\dag} \right],
\label{eq:R_int_main}
\end{align}
where the $(k,m)^{th}$ element of $\nbH^{(i)} \nbH^{(i)\dag} $ is
\begin{align}
\sum_{l=1}^{\Psi(n)} \left[\min\left\{1,\left(d^{(i)}\right)^{-\frac{\alpha}{2}}\right\} \right]^2 e^{-j2\pi d_{kl}^{(i)} \Psi(n)} e^{j2\pi d_{ml}^{(i)} \Psi(n)}.
\label{eq:HH_int_km}
\end{align}
Taking expectation in \eqref{eq:HH_int_km} with respect to the antenna locations of the interfering transmitters and using Lemma~\ref{lem:sh_uniform},  we have
\begin{align}
\nbbE[\nbR] \rightarrow \left(1 + \sum_{i \in \ncalI} P \left[\min\left\{1,\left(d^{(i)}\right)^{-\frac{\alpha}{2}}\right\} \right]^2 \right)\nbI
\label{eq:R_int_sum}
\end{align}
for large $\Psi(n)$.  Note that there are at most $8$ squares in the closest {\em ring} of interferers around the square in which the desired receiver 
is located, such that the total interference power contributed by this first ring of interferers to the summation term of~\eqref{eq:R_int_sum} is upper 
bounded by $8P$. Similarly, for $i>1$, there are at most $8i$ interferers in the $i^{th}$ ring, each at least at a distance $(i-1)c$, which upper bounds the interference contribution from the $i^{th}$ ring by $8i(c(i-1))^{-\alpha}$. This counting argument leads to the following upper bound on $\nbbE[\nbR]$
\begin{align}
\nbbE[\nbR] \leq \left(1 + 8P + 8c^{-\alpha} P \sum_{i=2}^{\infty} i (i-1)^{-\alpha} \right)\nbI,
\end{align}
that holds for sufficiently large $\Psi(n)$,  where the summation $\sum_{i=2}^{\infty} i (i-1)^{-\alpha} \leq \sum_{i=2}^{\infty} i^{1-\alpha}$ 
is convergent for $\alpha>2$.  Therefore, in the asymptotic regime of large $\Psi(n)$, have that $\nbbE[\nbR] \leq q\nbI$, for some constant 
$q$ independent of $n$. Substituting it back in \eqref{eq:Rshn}, we get
\begin{align}
\nbbE[R_{\rm sh} (n)] \geq \nbbE \left [ \log \det \left( \nbI + \frac{P}{q \Psi(n)} \nbH \nbH^{\dag} \right) \right ] \stackrel{(a)}{=} \Omega(\Psi(n)),
\end{align}
where $(a)$ follows from the spatial DoFs result given by \eqref{eq:DoF}. 
This leads to our second main result.

\begin{theorem}[Short hop: perturbed lattice] \label{thm:short-hop}
For the short hop strategy and the perturbed lattice model for the BS locations, $\Psi(n) = \sqrt{n}$ achieves $\nbbE[R_{\rm sh} (n)] = \Omega(\sqrt{n})$ for each short hop, and hence ergodic rate of $\Omega(1)$ for each source-destination pair, where expectation is over antenna locations.
\hfill $\blacksquare$
\end{theorem}

\begin{remark}[Scalability] \label{rem:scalability}
Comparing Theorems~\ref{thm:long-hop} and~\ref{thm:short-hop}, we note that short hop strategy is significantly better for network scalability. While for any $\alpha>2$ it requires only $\Psi(n) = \sqrt{n}$, the antenna requirement in long hop is always higher and keeps increasing further with $\alpha$.
\end{remark}

Note that Theorem~\ref{thm:short-hop} would be enough to claim achievability of the scaling law 
if the BSs were allowed to randomly select their antenna locations independently and uniformly in their allowed fixed area, 
such that the expectation of~\eqref{eq:Rshn} has the operational meaning of ``ergodic rate'', achieved by coding over a long sequence of 
realizations of the antenna locations. 
A more practical viewpoint is to consider the achievability of the scaling law with high probability for a fixed random realization of the antennas. For such a result, 
we need to show that $\nbbP(R_{\rm sh} (n) \geq r \Psi(n)) \rightarrow 1$ as $n \rightarrow \infty$, for some constant $r > 0$. 
One way to establish this result is by showing that $R_{\rm sh} (n)$ concentrates around its mean,  
in the sense that $\nbbP(|R_{\rm sh} (n) - \nbbE[R_{\rm sh} (n)]| > \epsilon \Psi(n)) \rightarrow 0$, with $\epsilon>0$, as $n \rightarrow \infty$. Such concentration result is shown in \cite[Lemma 2.2]{OzgLevJ2013} for the case of an isolated (without interference) LoS MIMO link. 
In Section~\ref{sec:Discussion}, we provide numerical evidence of this concentration. 
Complementary to this and the achievability proof in~\cite[Lemma 2]{LeeChuJ2012}, we propose a simpler approach 
in Appendix~\ref{app:achievability} based on the spectral radius results of \cite{DesLevC2013b}. Note that Appendix~\ref{app:achievability} merely suggests a direction for future work and should not be interpreted as a formal result due to two reasons: (i) we use a slightly tighter bound on spectral radius than \cite{DesLevC2013b} to illustrate the approach (the tighter bound needs proof), and (ii) the bound in \cite{DesLevC2013b} is itself derived under an element-wise approximation of the LoS channel matrix by another random matrix, which needs to be rigorized.

\subsection{Throughput for the Random Network Model}

\begin{figure}
\centering
\includegraphics[width=\columnwidth]{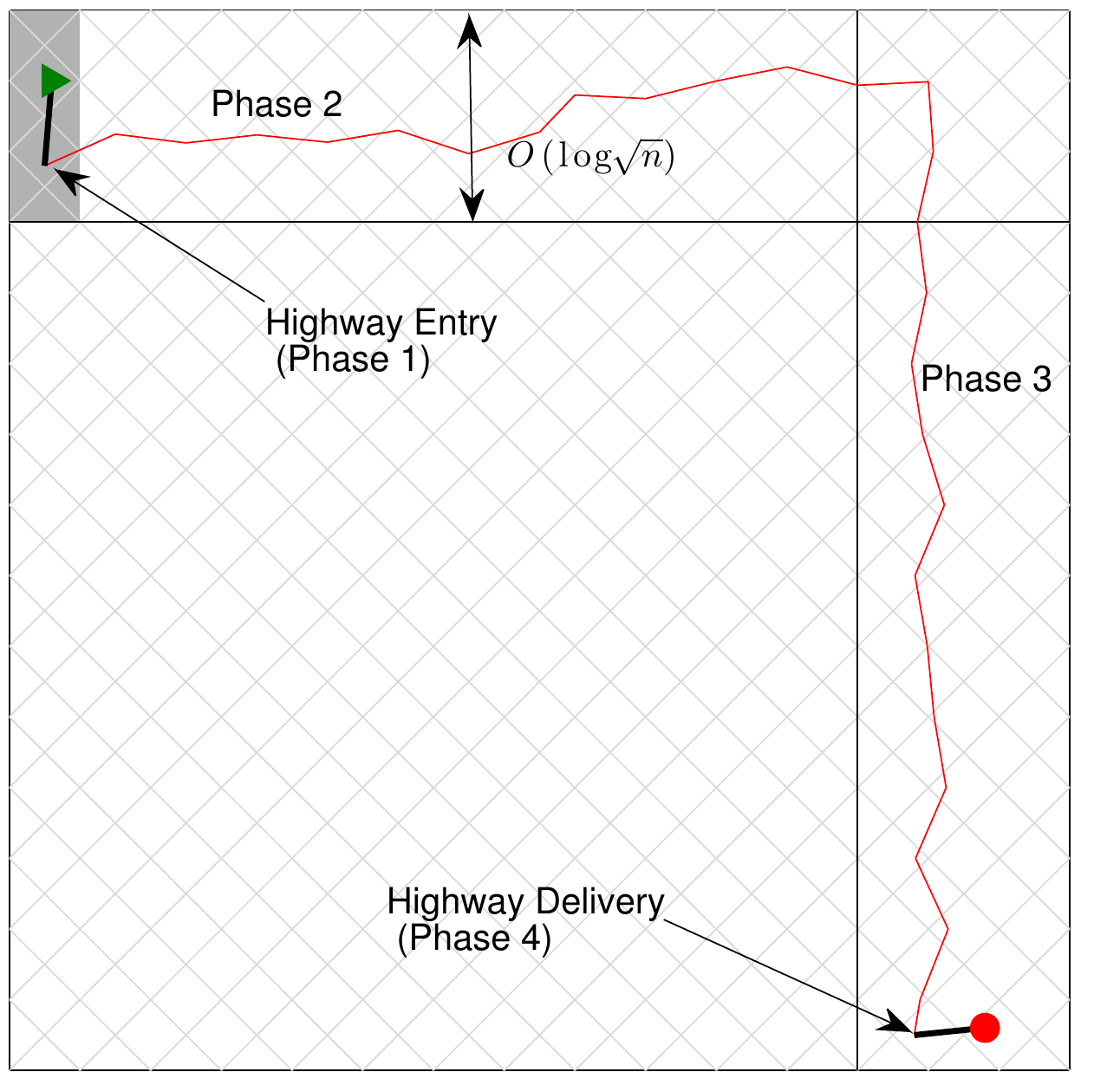}
\caption{A schematic sketch of 4-phase routing strategy for a random network. The highway entry (phase 1) and exit (phase 4) are denoted by thicker lines. The triangle and circle denote source and destination BSs, respectively.}
\label{fig:Percolation}
\end{figure}

It is easy to generalize the above achievability result to the 
random network model using percolation theory~\cite{FraDouJ2007}. 
The key ideas are explained with the help of  Fig.~\ref{fig:Percolation}. 
The source and the destination BSs of a given pair are denoted by a triangle and circle, respectively. The routing protocol is formed by 
four phases. In the first phase, the source BS sends its data to a {\em horizontal highway}. 
In the second phase, the data is communicated over a horizontal highway, until it reaches the {\em vertical highway} 
associated with the destination BS. In the third phase, data is transmitted over this vertical highway. Finally, in the fourth phase,
the data is delivered to the destination by the last hop off the highway.

In~\cite{FraDouJ2007}, it is shown that w.h.p. there are $\Omega(\sqrt{n})$ {\em horizontal highways}, connecting the left and right edges of $B_n$ such that the distance between any two neighboring BSs on each highway is bounded by a constant independent of $n$. Similarly there are $\Omega(\sqrt{n})$ highways connecting the top and bottom edges. It is further shown in~\cite{FraDouJ2007} that these highways are distributed uniformly over the network. In particular, these horizontal and vertical highways can be respectively grouped into disjoint groups of $\log(\sqrt{n})$ highways, with each group confined in a rectangular slab of size $\sqrt{n}\times \log \sqrt{n}$. One each of such horizontal and vertical slabs is also illustrated in Fig.~\ref{fig:Percolation}. Due to this regularity property, it is shown in~\cite{FraDouJ2007} that it is possible to uniquely 
associate a slab of width $O(1)$ with each highway, such that each highway carries data of $O(\sqrt{n})$ source-destination pairs. 
Hence, phases 2 and 3 of the above routing protocol are equivalent to what was already discussed in the context of perturbed lattice model. 
In particular, with $\Psi(n) =\sqrt{n}$ it is possible for each highway to simultaneously achieve rate $\Omega(\sqrt{n})$, where it should be noted 
that the interference power can be bounded in exactly the same way as demonstrated in the proof of Theorem~\ref{thm:short-hop}. 
This rate is then equally distributed over $O(\sqrt{n})$ connections, such that each achieves rate $\Omega(1)$. 
which leads to the above discussed properties. Interested readers can refer to~\cite{FraDouJ2007} for more details about this construction. 

Now, we need to show that phases 1 and 4 are not the rate bottlenecks.
The argument is the same for both the phases, so we explain it in terms of phase 1. The width of the horizontal and vertical slabs yields 
that the distance between a source and the nearest highway node is $O(\log \sqrt{n})$. 
Substituting this distance in \eqref{eq:DoF} and recalling 
that $\lambda \propto \Psi(n)^{-1}$, it is easy to see that the spatial DoFs for this channel are $\Omega\left(\frac{\sqrt{n}}{\log\sqrt{n}}\right)$ and that
the interference power can be bounded in the same way as done in the previous subsection. 
To complete the argument, we have to find the number of BSs sharing the same highway entry point. 
The rate $\Omega\left(\frac{\sqrt{n}}{\log\sqrt{n}}\right)$ must be shared amongst these BSs. 
As shown in Fig.~\ref{fig:Percolation}, these source BSs are confined 
in the gray region of size $O(1)\times O(\log \sqrt{n})$. By Lemma~\ref{lem:Chernoff_Poisson}, the number of BSs in this region 
is upper bounded by  $O(\log \sqrt{n})$. Therefore, the rate per source node in phase 1 is lower bounded by 
$\Omega\left(\frac{\sqrt{n}}{(\log \sqrt{n})^2}\right)$, which shows that phases 1 and 4 are not the rate bottleneck of the system. 

\begin{figure}
\centering
\includegraphics[width=.49\columnwidth]{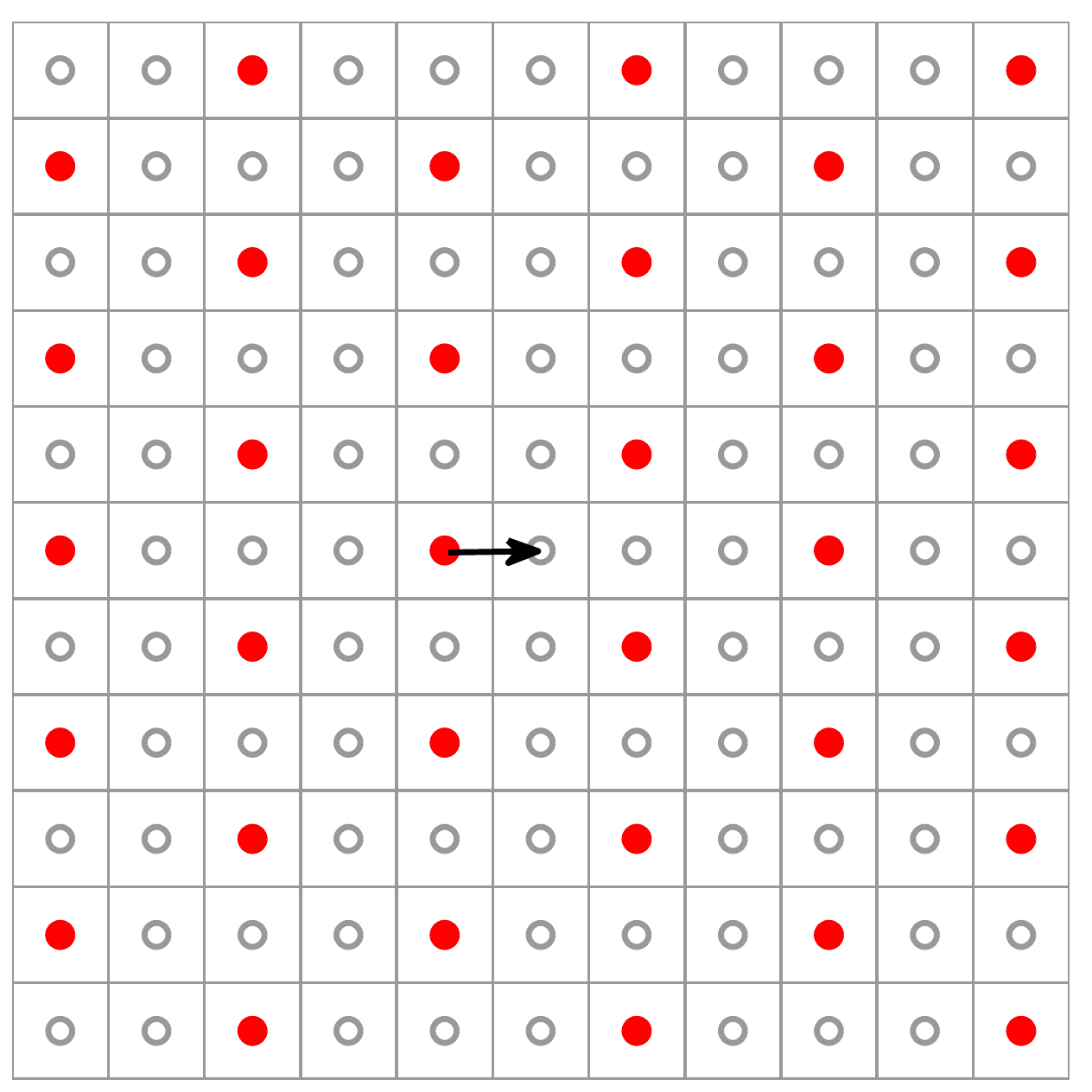}
\includegraphics[width=.49\columnwidth]{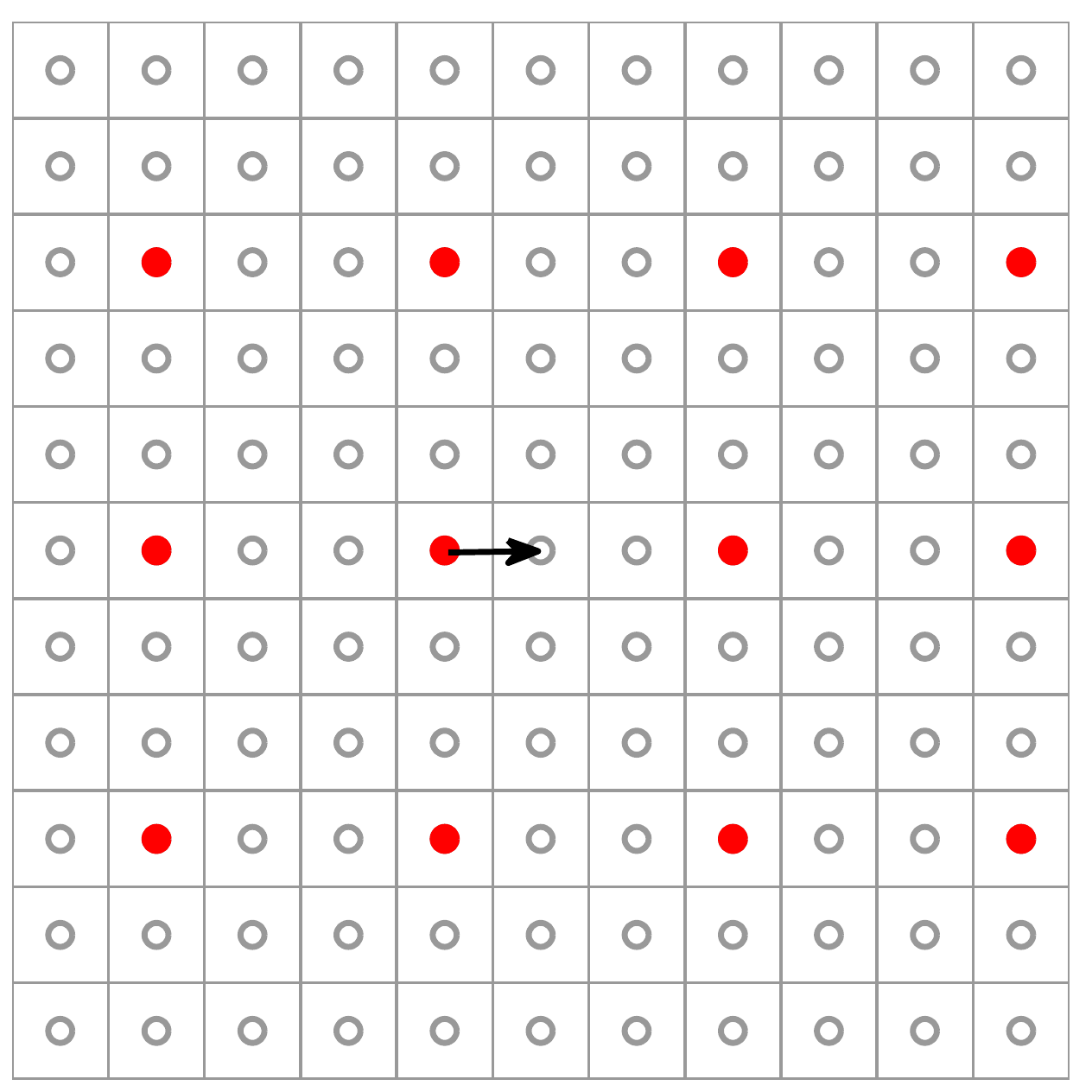}
\caption{Deterministic reuse patterns: {\em(first)} 4-reuse, and {\em (second)} 9-reuse. The BSs active in the current time slot are denoted by solid circles. The LoS MIMO link under investigation is denoted by a one-sided arrow.}
\label{fig:ReusePatterns}
\end{figure}

\section{Discussion} \label{sec:Discussion}

In this section we discuss some additional aspects and consequences of throughout scaling analysis of wireless backhaul networks 
carried out in the previous sections.

\paragraph{Numerical Evaluation of Achievable Rates}
In Section~\ref{sec:ShortHop} we showed that it is possible to implement a scalable wireless backhaul network using short hop strategy, even in LoS propagation. In this section, we take a step forward and find a ball-park number for the rate achievable by a short hop link.
For simplicity, we assume that the BSs are located on a squared lattice. 
To minimize edge effects, we simulate a lattice with $529$ BSs, with the desired receiver at the lattice center (see Fig.~\ref{fig:ReusePatterns} where
the short hop link of interest is denoted by a one-sided arrow). 
We assume carrier frequency of $30$GHz and pathloss exponent $\alpha=5$.\footnote{Note that the characterization of pathloss for mm-wave communication 
is currently under active investigation (see \cite{MacZhaC2013}  for some recent measurement results).}
Letting $\mu = \frac{Pc^{-\alpha}}{N_0 W}$ denote  the SNR for the desired link under single antenna transmission 
in the absence of interference, we set the desired reference value $\mu = 0$ dB, which means that such single antenna link in isolation would achieve 
a rate of $\log_2(1+\mu) = 1$ bps/Hz. 
For the LoS MIMO link, we use the same parameters as in Remark~\ref{rem:urban}, with inter-BS distance of $100$m,
and  $\Psi(n)=64$ BS antennas uniformly distributed in a square of side $8$m.
 The achievable rate for the desired link can be numerically computed using~\eqref{eq:Rshn}. Assuming 
that all the BSs transmit simultaneously (i.e., {\em full-reuse}), the rate is about $17$ bps/Hz, which means that with only 
about $100$ MHz bandwidth (well within what it is anticipated in the 30 GHz band), 
a $1.7$ Gbps backhaul link can be implemented with fairly standard technology. 

\begin{figure}
\centering
\includegraphics[width=\columnwidth]{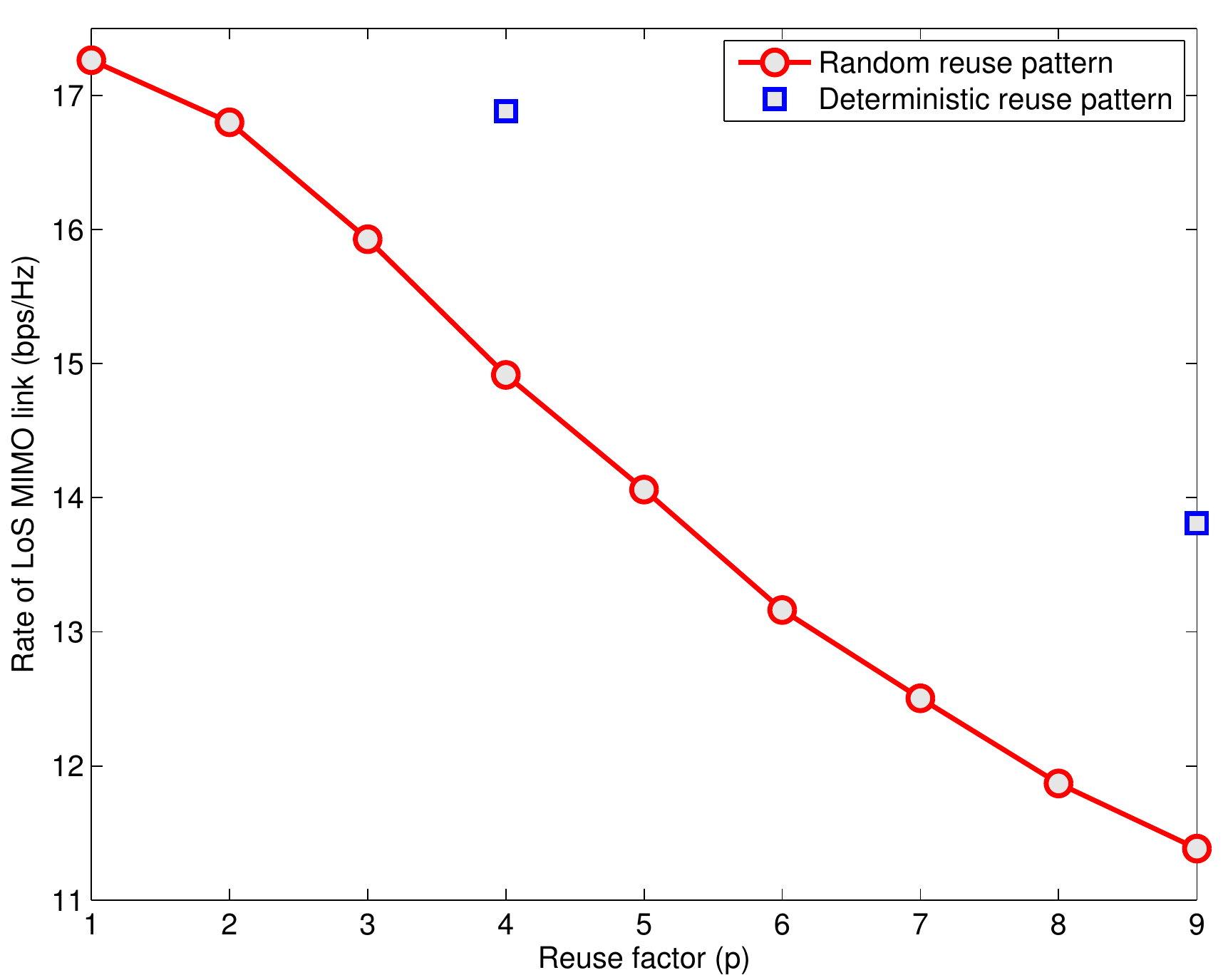}
\caption{Rate of LoS MIMO link as a function of the reuse factor. In addition to the random reuse, the deterministic reuse patterns of Fig.~\ref{fig:ReusePatterns} are also simulated.}
\label{fig:ReusePlot}
\end{figure}

\paragraph{Effect of Time/Frequency Reuse}
As is usually done in cellular systems, it is possible to introduce a resource reuse factor $p>1$ in a wireless backhaul system as well, i.e., the network is partitioned into $p$ sub-networks, each of which is active on a fraction $1/p$ of the total transmission resource (time-frequency slots). 
Fig.~\ref{fig:ReusePatterns} shows two examples for a square lattice network with $p = 4$ and $p = 9$ reuse factors. While it was possible to partition the BSs into $p$ meaningful sets in these two cases, it is not always easy to accomplish this for any given $p$. Therefore, we also consider a {\em random reuse} strategy where we randomly select with uniform probability $1/p$ 
the BSs using the same resource. In all cases, the link rate with reuse $p$ is given by
\begin{align}
R_{\rm sh} (n) \geq \frac{1}{p}\log \det \left( \nbI + \frac{pP}{\Psi(n)} \nbR^{-1} \nbH \nbH^{\dag} \right),
\label{eq:Rshn_p_reuse}
\end{align}
where $\nbR$ now includes interference from only those BSs that are active on the same resource partition as the desired link, and the transmit power is increased 
by the factor $p$ since each BS is active only on a fraction $1/p$ of the channel slots.  For the same simulation setup as the previous subsection, 
we numerically evaluate~\eqref{eq:Rshn_p_reuse} for various values of $p$ and plot the mean rates in Fig.~\ref{fig:ReusePlot}. Two remarks are in order: (i) for random reuse, $p = 1$ (universal reuse) is the best option for this set of simulation parameters; and (ii) deterministic (planned) reuse performs significantly better both for $p=4$ and $9$.

\begin{figure}
\centering
\includegraphics[width=\columnwidth]{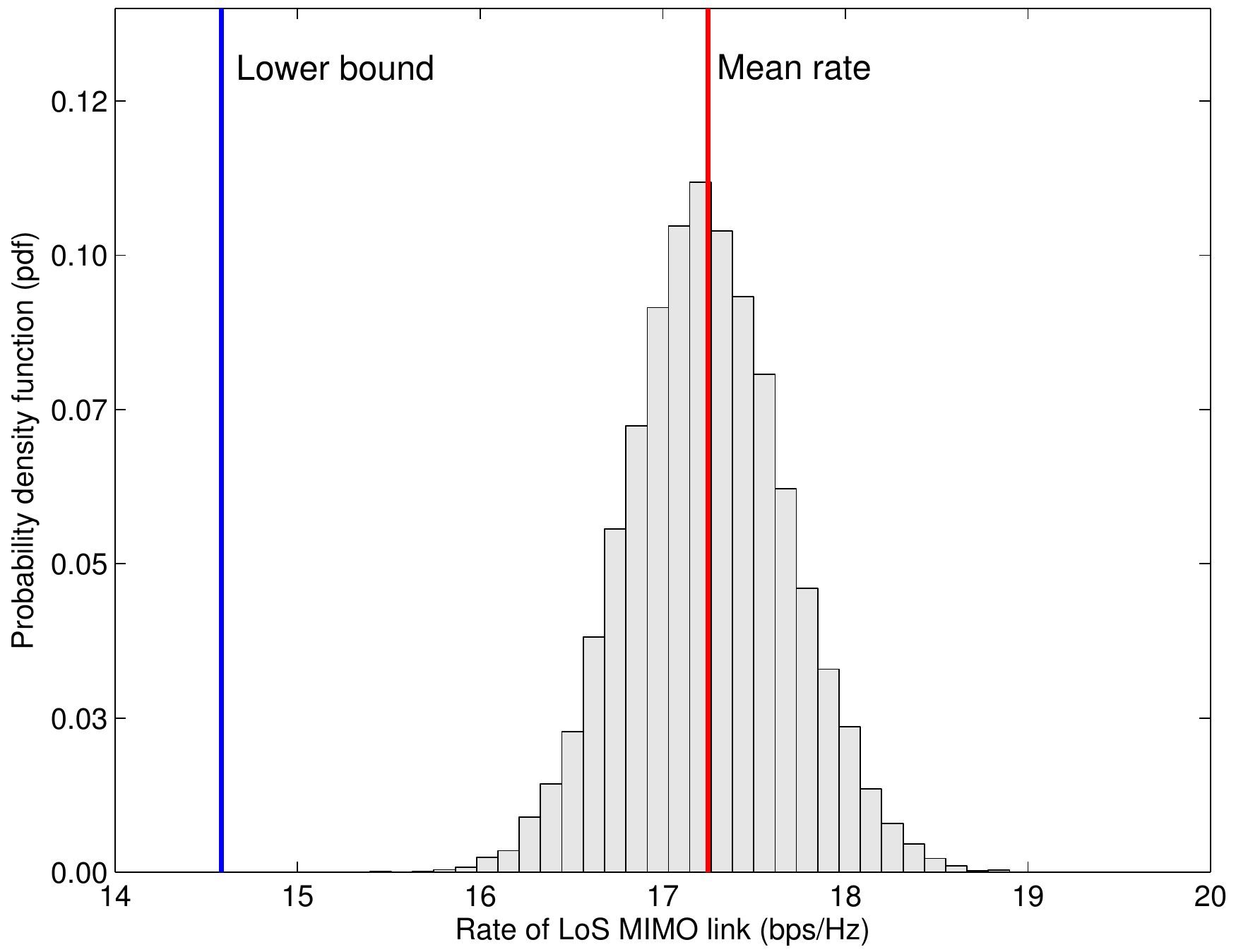}
\includegraphics[width=\columnwidth]{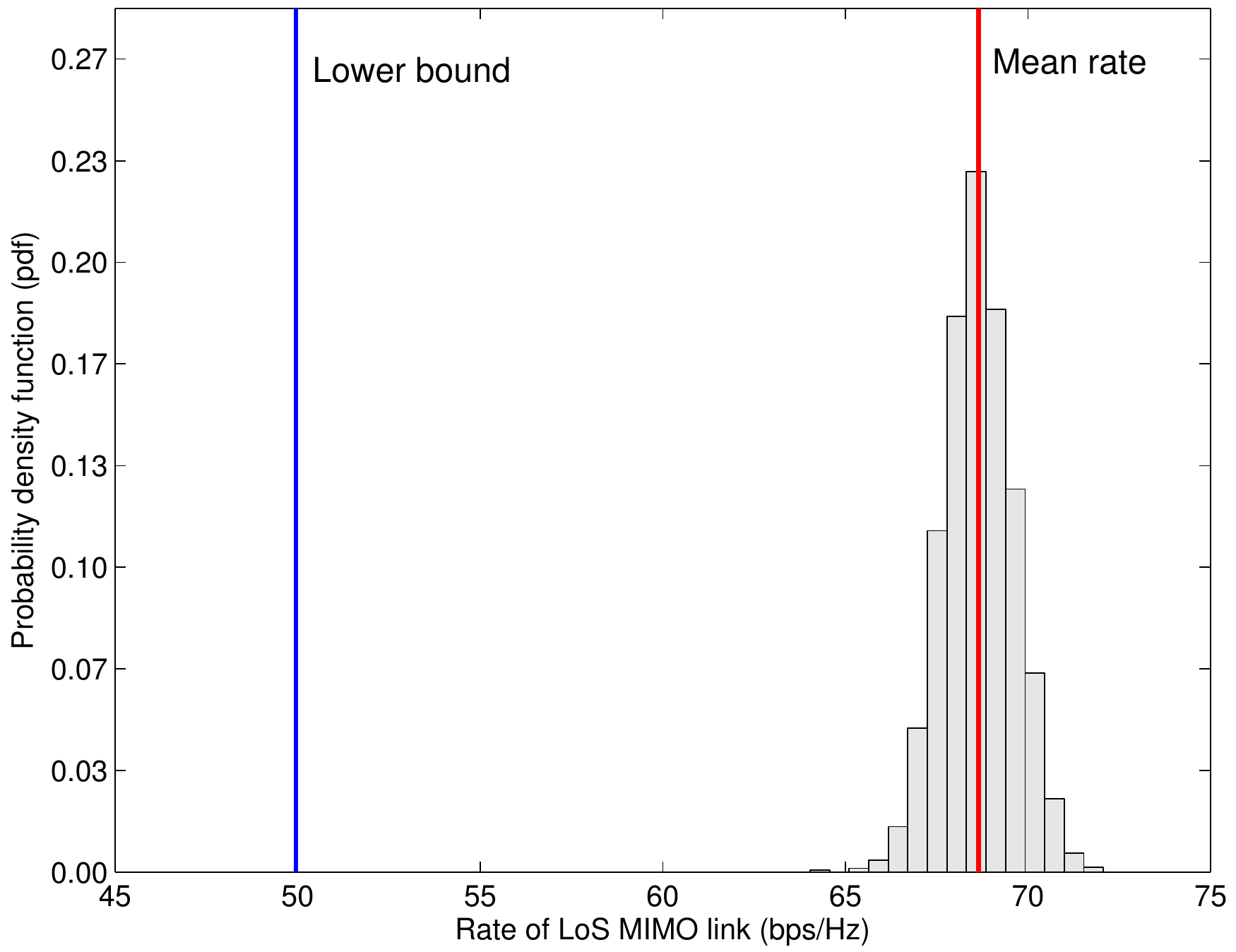}
\caption{Probability density function (pdf) of the LoS MIMO link rate. {\em (first)} 64 antennas, and {\em (second)} 256 antennas. Also plotted are the mean rate and the lower bound on the mean rate.}
\label{fig:Ratepdf}
\end{figure}

\paragraph{Distribution of the rate of LoS MIMO Link} \label{sec:Rateconcentration}
While the previous two subsections focused on the mean rates achievable by a LoS MIMO link, in this subsection we explore the link rate distribution.
We consider the same setup as the previous two subsections and evaluate the achievable rate using~\eqref{eq:Rshn} with full reuse. Instead of averaging 
with respect to the antenna placement, we plot the rate distributions in Fig.~\ref{fig:Ratepdf} for two cases: (i) $\Psi(n) = 64$, and (ii) $\Psi(n) = 256$.
We also plot the mean rate obtained by Monte Carlo simulation and the ``ergodic'' lower bound on the mean rate given by~\eqref{eq:RshnExp}. 
Note that the rate distribution is fairly concentrated around its mean, and this concentration is more and more evident 
when the number of antennas increases. Note also that the ergodic rate lower bound developed and analyzed in this paper is concisely on the left of the 
actual rate distribution.  As discussed in Section \ref{sec:ShortHop_Lattice} after Remark \ref{rem:scalability}, this indicates that
the actual achievable rate for a random placement of the BS antennas has the same scaling as the ergodic lower bound w.h.p. (a stronger achievability conclusion than Theorem~\ref{thm:short-hop}). See also Appendix \ref{app:achievability} for more discussion on this point.

\paragraph{Availability and Impact of Wired Backhaul}
Throughout this paper, we assumed that none of the BSs has access to wired backhaul. This can be easily relaxed in some cases of practical relevance. For instance, let $B_n$ represent an urban cellular network with some of the BSs (gateways) at the boundary (perimeter) having access to wired backhaul. It is reasonable to assume that each cell within $B_n$ generates fraction $\rho$ of traffic with destinations inside $B_n$ and a fraction $1-\rho$ of traffic with destinations outside the network. Therefore, for any $\rho \in (0,1)$, $\Theta(n)$ traffic must flow through the gateways into the wired backbone network. Following short hop strategy with $\Psi(n) = \sqrt{n}$, each gateway BS can relay $\Omega(\sqrt{n})$ traffic to the ``outside world'', which means we need $O(\sqrt{n})$ gateways at the boundary of $B_n$ to relay a total of $\Theta(n)$ traffic.

Another setup of interest is when $O(n^{\beta}), \beta \in [0,1]$ BSs inside $B_n$ have access to high-capacity wired backhaul. 
For the ease of exposition, let us consider the lattice model with regular placement of the wired backhaul BSs. Then, this setup reduces to $n^{\beta}$ mini-networks 
of $n^{1-\beta}$ BSs each. By Theorem~\ref{thm:short-hop}, each BS needs $\Psi(n) = \Omega(n^{\frac{1-\beta}{2}})$ antennas to make each one of these mini-networks scalable. By bounding the interference power as in the proof of Theorem~\ref{thm:short-hop}, it is easy to show that the mini-networks can operate simultaneously, thereby achieving scalability of the whole network. A tiny sample of related works in this direction is~\cite{ZemdeVJ2005,ShiJeoJ2011}.

\section{Concluding remarks}

Wireless backhaul for current cellular networks is quickly becoming a necessity, especially in the context of urban small cell deployments. Two likely features of future backhaul networks are: (i) higher transmission frequencies (mm-waves), and consequently, (ii) large number of transmit antennas per BS. 
We modeled this network as a multi-antenna random extended network, where the number of antennas per BS can scale as some arbitrary function 
of the total number of BSs. Using geometric arguments, we derived an information theoretic upper bound on the capacity of this network, and consequently, a lower bound on the number of antennas per BS required for throughput scalability. We further compared the scalability requirements of two competing strategies of interest. While the first minimizes the number of hops by forming thin beams and ideally achieving full beamforming gain over each hop, the other communicates data from source to its destination through many short hops, each achieving full multiplexing gain. Although it may seem intuitive at first to minimize the number of hops by forming narrow beams and hence minimizing interference to neighboring nodes, we show that the short hop strategy is significantly better. The achievability result for the short hop strategy is derived under LoS propagation while carefully accounting for the the fundamental limits on the spatial DoFs of a LoS MIMO channel. 

From the practical system design viewpoint, some important guidelines can be drawn from our analysis. 
In the large-scale MIMO regime advocated by the short-hop architecture, large spectral efficiency per link can be achieved with very simple coding and modulation schemes. For example, in the numerical calculations of Section \ref{sec:Discussion} we achieve
$17$ bps/Hz with $64$ antennas per BS, which implies $0.266$ bit/stream. This is easily achievable by using QPSK concatenated with power 
capacity-approaching binary (Turbo/LDPC) codes. 
Reuse results in Fig.~\ref{fig:ReusePlot} highlight the importance of well-planned reuse, which along with the {\em optimal} distributed power control forms an important area of future work under the umbrella of self-organizing backhaul networks. 
Finally, the simple extension of our analysis to the case where some  BSs have access to wired backhaul (Section \ref{sec:Discussion}) 
yields an appealing system design tradeoff:  
the per-BS hardware complexity (number of antennas) can be reduced from $n^{1/2}$ to $n^{(1 - \beta)/2}$ in the presence of
$n^\beta$ wired backhaul BSs uniformly placed in the network. 

Concrete directions for future work may include: (i) develop a communication theoretic channel model that captures the DoFs bottleneck of electromagnetic propagation \cite{FraMigJ2009} without necessarily assuming LoS; (ii) rigorize the approximation argument in \cite{DesLevC2013,DesLevC2013b} which is instrumental for the sharp characterization of the DoFs of LoS MIMO channels; and (iii) performance evaluation of backhaul wireless networks under realistic population conditions in order to assess if the guidelines obtained from scaling laws reflect in actual system gains. 

\appendix

\subsection{Proof of Lemma~\ref{lem:Chernoff_Poisson}} \label{app:Chernoff_Poisson}
Recall that for $X\sim \Pois(\lambda_{\rm b})$, we have the following Chernoff bound for all $x>\lambda_{\rm b}$
\begin{align}
\nbbP(X \geq x) \leq \exp(-\lambda_{\rm b}) \left(\frac{\exp(1) \lambda_{\rm b}}{x} \right)^x.
\end{align}
Please refer to~\cite[Appendix II]{FraDouJ2007} or~\cite[Theorem 5.4]{MitUpfB2005} for further details. Also note that $N(\ncalA) \sim \Pois(\lambda_{\rm b} |\ncalA|)$~\cite{BacBlaB2009}. Specializing the above bound for $N(\ncalA)$ and substituting $x = 2 \lambda_{\rm b} |\ncalA|$, we get
\begin{align}
\nbbP(N(\ncalA)\geq 2 \lambda_{\rm b} |\ncalA|) &\leq e^{-\lambda_{\rm b} |\ncalA|} \left( \frac{e}{2}\right)^{2\lambda_{\rm b} |\ncalA|} = \left( \frac{e}{4}\right)^{\lambda_{\rm b} |\ncalA|},
\end{align}
from which~\eqref{eq:Chernoff_Poisson_UB} follows by the fact that $\frac{e}{4} < 1$. Similarly, for all $x<\lambda_{\rm b}$, we have the same Chernoff bound
\begin{align}
\nbbP(X \leq x) \leq \exp(-\lambda_{\rm b}) \left(\frac{\exp(1) \lambda_{\rm b}}{x} \right)^x.
\end{align}
Again specializing this bound for $N(\ncalA)$ and substituting $x =  \frac{\lambda_{\rm b}}{2} |\ncalA|$, we get
\begin{align}
\nbbP\left(N(\ncalA)\leq \frac{\lambda_{\rm b}}{2} |\ncalA|\right) &\leq e^{-\lambda_{\rm b} |\ncalA|} (2e)^{\frac{\lambda_{\rm b}}{2} |\ncalA|} = \left(\frac{2}{e} \right)^{\frac{\lambda_{\rm b}}{2} |\ncalA|},
\end{align}
from which~\eqref{eq:Chernoff_Poisson_LB} follows by the fact that $\frac{2}{e} < 1$. \hfill 
\IEEEQED

\subsection{Proof of Lemma~\ref{lem:Cs_UB}} \label{app:Cs_UB}
For notational simplicity, define  $\kappa = \frac{\alpha}{2} - 2- \log_n \Psi(n)$,
and assume it to be positive, which implies that this derivation is applicable for 
$\alpha > 2 \left(2 + \log_n \Psi(n) \right)$.
This condition will be required for the Taylor expansion of the log terms. 
The summation term $C_{\rm s}$ can be expressed as
\begin{align}
C_{\rm s} &= \sum_{i=1}^{ \log \frac{\sqrt{n}}{2}} \frac{n} {e^i}  \log \left( 1 + P n^{2-\frac{\alpha}{2}} \Psi(n)2^{\alpha} e^{i\alpha} \right) \nonumber\\
&= \sum_{i=1}^{ \log \frac{\sqrt{n}}{2}} \frac{n} {e^i}  \log \left( 1 + P 2^{\alpha} \frac{e^{i\alpha}}{n^{\kappa}} \right) \nonumber\\
&= \sum_{i=1}^{ \frac{2\kappa}{\alpha} \log \frac{\sqrt{n}}{2}} \frac{n} {e^i}  \log \left( 1 + P 2^{\alpha} \frac{e^{i\alpha}}{n^{\kappa}} \right) + \nonumber \\
&\sum_{i=\frac{2\kappa}{\alpha} \log \frac{\sqrt{n}}{2} + 1}^{ \log \frac{\sqrt{n}}{2}} \frac{n} {e^i}  \log \left( 1 + P 2^{\alpha} \frac{e^{i\alpha}}{n^{\kappa}} \right)
= C_{\rm s_1} + C_{\rm s_2}.
\label{eq:C_S12}
\end{align}
Since the constant $P2^{\alpha}$ in the log terms of both $C_{\rm s_1}$ and $C_{\rm s_2}$ is independent of $n$ and hence does not impact scaling of these terms, we will ignore it in the following discussion for notational simplicity. 
Using Taylor series expansion for log term, $C_{\rm s_1}$ can now be expressed as
\begin{align}
C_{\rm s_1} &= \sum_{i=1}^{ \frac{2\kappa}{\alpha} \log \frac{\sqrt{n}}{2}} \frac{n} {e^i} \sum_{k=1}^\infty \frac{(-1)^{k+1}}{k} \frac{e^{ki\alpha}}{n^{k\kappa}} \nonumber \\
&= n\sum_{k=1}^\infty \frac{(-1)^{k+1}}{k} \frac{1}{n^{k\kappa}} \sum_{i=1}^{ \frac{2\kappa}{\alpha} \log \frac{\sqrt{n}}{2}} e^{i(k\alpha - 1)}.
\label{eq:CS1_1}
\end{align}
The summation with respect to $i$ can be computed as
\begin{align}
\sum_{i=1}^{ \frac{2\kappa}{\alpha} \log \frac{\sqrt{n}}{2}} e^{i(k\alpha - 1)} &= \frac{e^{k\alpha - 1}}{e^{k\alpha - 1} - 1} \left(\frac{n^{(k\alpha-1)\kappa/\alpha}}{2^{2(k\alpha-1)\kappa/\alpha}} -1\right) \nonumber \\  
&\stackrel{(a)}{\leq} m \left(\frac{n^{(k\alpha-1)\kappa/\alpha}}{2^{2(k\alpha-1)\kappa/\alpha}} -1\right),
\label{eq:CS1_sum_int1}
\end{align}
where $(a)$ follows by the fact that $\frac{e^{k\alpha - 1}}{e^{k\alpha - 1} - 1}$ is uniformly upper bounded by some positive constant $m$. Substituting \eqref{eq:CS1_sum_int1} back in \eqref{eq:CS1_1}, we get
\begin{align}
C_{\rm s_1} &\leq n\sum_{k=1}^\infty \frac{(-1)^{k+1}}{k} \frac{1}{n^{k\kappa}} m \left(\frac{n^{(k\alpha-1)\kappa/\alpha}}{2^{2(k\alpha-1)\kappa/\alpha}} -1\right) \nonumber \\
&= C_{\rm s_{11}} + C_{\rm s_{12}}.
\label{eq:CS1_sum_int2}
\end{align}
Ignoring again the constants, $C_{\rm s_{11}}$ can be upper bounded as
\begin{align}
C_{\rm s_{11}} &= n^{1-\frac{\kappa}{\alpha}} \sum_{k=1}^\infty \frac{(-1)^{k+1}}{k} = n^{1-\frac{\kappa}{\alpha}} \log 2
= O(n^{1-\frac{\kappa}{\alpha}}).
\label{eq:CS11}
\end{align}
Similarly, $C_{\rm s_{12}}$ can be upper bounded as  
\begin{align}
C_{\rm s_{12}} & = n \sum_{k=1}^\infty \frac{(-1)^{k+1}}{k} \frac{1}{n^{k\kappa}} = n \log(1 + n^{-\kappa})  = O(n^{1-\kappa}). 
\label{eq:CS12}
\end{align}
Substituting \eqref{eq:CS11} and \eqref{eq:CS12} back in \eqref{eq:CS1_sum_int2}, we get 
\begin{align}
C_{\rm s_1} \leq O(n^{1-\frac{\kappa}{\alpha}}) + O(n^{1-\kappa}) = O(n^{1-\frac{\kappa}{\alpha}}),
\label{eq:CS1_final}
\end{align}
where the last equality follows from the fact that $\alpha>2$. We now turn our attention to $C_{\rm s_2}$ in~\eqref{eq:C_S12}, where we again ignore the constants. It can be expressed as
\begin{align}
C_{\rm s_2} &= \sum_{i=\frac{2\kappa}{\alpha} \log \frac{\sqrt{n}}{2} + 1}^{ \log \frac{\sqrt{n}}{2}} \frac{n} {e^i}  \log \left( 1 + \frac{e^{i\alpha}}{n^{\kappa}} \right) \nonumber \\
&= \sum \frac{n} {e^i} \log \left(e^{i\alpha} \cdot \frac{1}{n^\kappa} \cdot \left(1 + \frac{n^\kappa}{e^{i\alpha}} \right) \right) \nonumber \\
&= \sum \frac{n\alpha i}{e^i} - \sum \frac{n\kappa}{e^i} \log n + \sum \frac{n} {e^i} \log \left(1 + \frac{n^\kappa}{e^{i\alpha}} \right) \nonumber \\
&= C_{\rm s_{21}} - C_{\rm s_{22}} + C_{\rm s_{23}},
\label{eq:CS2_f_int}
\end{align}
where the limits of the summation for all the terms are the same as the first equation. We now look at the three terms separately starting with $C_{\rm s_{21}}$
\begin{align}
C_{\rm s_{21}} &= n \alpha \sum_{i=\frac{2\kappa}{\alpha} \log \frac{\sqrt{n}}{2} + 1}^{ \log \frac{\sqrt{n}}{2}} \frac{i}{e^i} \nonumber \\
&\stackrel{(a)}{\leq} O(\sqrt{n}) + O\left(n^{1-\frac{\kappa}{\alpha}}\right) \stackrel{(b)}{=} O\left(n^{1-\frac{\kappa}{\alpha}}\right),
\label{eq:CS21_f}
\end{align}
where $(a)$ follows by computing the summation directly, and $(b)$ from the fact that $\frac{\kappa}{\alpha} = \frac{1}{2} - \frac{1}{\alpha}\left(2+ \log_n \Psi(n) \right) \leq \frac{1}{2}$. Similarly, the second term can be expressed as
\begin{align}
C_{\rm s_{22}} &= n \kappa \log(n) \sum_{i=\frac{2\kappa}{\alpha} \log \frac{\sqrt{n}}{2} + 1}^{ \log \frac{\sqrt{n}}{2}} e^{-i} 
\stackrel{(a)}{\leq} O(\sqrt{n}\log(n)) + \nonumber \\
&O\left(n^{1-\frac{\kappa}{\alpha} } \log(n)\right)
= O\left(n^{1-\frac{\kappa}{\alpha} } \log(n)\right),
\label{eq:CS22_f}
\end{align}
where $(a)$ again follows by computing the summation directly and the final result by the fact that $\kappa < \frac{\alpha}{2}$. Now we come to the final term $C_{\rm s_{23}}$, which can be expressed in terms of the Taylor series, $\log(1+x) = \sum_{k=1}^{\infty} \frac{(-1)^{k+1}}{k}x^k, |x| \leq 1,$ as
\begin{align}
C_{\rm s_{23}} &= n \sum_{i=\frac{2\kappa}{\alpha} \log \frac{\sqrt{n}}{2} + 1}^{ \log \frac{\sqrt{n}}{2}} \frac{1}{e^{i}} \log \left(1 + \frac{n^\kappa}{e^{i\alpha}} \right) \nonumber \\
&= n \sum_{k=1}^\infty \frac{(-1)^{k+1}}{k} n^{k\kappa} \sum_{i=\frac{2\kappa}{\alpha} \log \frac{\sqrt{n}}{2} + 1}^{ \log \frac{\sqrt{n}}{2}} \frac{1}{e^{i(k\alpha+1)}}.
\end{align}
Note that the above expansion of $\log$ term only holds when $\frac{n^\kappa}{e^{i\alpha}}<1$. Since $\frac{n^\kappa}{e^{i\alpha}}$ is a decreasing function of $i$, it is sufficient to show that the condition holds for $i = \frac{2\kappa}{\alpha} \log \frac{\sqrt{n}}{2} + 1$. For this value of $i$, it is easy to show that $\frac{n^\kappa}{e^{i\alpha}} = \frac{2^{2\kappa}}{e^\alpha} \stackrel{(a)}{<} \left(\frac{2}{e} \right)^\alpha \stackrel{(b)}{<} 1$, where $(a)$ follows from the fact that $2\kappa < \alpha$, and $(b)$ is due to $\frac{2}{e} < 1$ and $\alpha>2$. Now, the summation with respect to $i$ can be expressed as
{\small
\begin{align}
&\frac{e^{-k\alpha - 1}}{1-e^{-k\alpha-1}} \frac{1}{e^{\frac{k\alpha + 1}{\alpha} 2\kappa \log \frac{\sqrt{n}}{2}}} \left[1 - \frac{e^{k\alpha+1}} {e^{(k\alpha+1) \left(1-\frac{2\kappa}{\alpha} \right) \log \frac{\sqrt{n}}{2}}} \right] \nonumber \\
&= \frac{e^{-k\alpha - 1}}{1-e^{-k\alpha-1}} \left(\frac{4}{n} \right)^{k\kappa + \frac{\kappa}{\alpha}}  - \frac{1}{1-e^{-k\alpha - 1}} 
\left(\frac{4}{n} \right)^{\frac{\alpha k + 1}{2}}.
\end{align}
}
Using the fact that both $\frac{e^{-k\alpha - 1}}{1-e^{-k\alpha-1}}$ and $\frac{1}{1-e^{-k\alpha - 1}}$ are upper bounded uniformly by positive constants, and ignoring the constants that do not impact scaling laws, $C_{\rm s_{23}}$ can be upper bounded by the sum of the following two terms
\begin{align}
C_{\rm s_{231}}  &= n^{1-\frac{\kappa}{\alpha}} \sum_{k=1}^\infty \frac{(-1)^{k+1}}{k} = n^{1-\frac{\kappa}{\alpha}} \log 2 = O\left( n^{1-\frac{\kappa}{\alpha}} \right)\\
C_{\rm s_{232}} &= \sqrt{n} \sum_{k=1}^\infty \frac{(-1)^{k+1}}{k} n^{(\kappa - \frac{\alpha}{2})k} \stackrel{(a)}{=} \sqrt{n} \log(1 + n^{\kappa - \frac{\alpha}{2}}) \nonumber \\
&= O\left(\frac{\sqrt{n}}{n^{\frac{\alpha}{2} - \kappa}} \right) = O(\sqrt{n}),
\end{align}
where $(a)$ follows from the fact that since $\kappa - \frac{\alpha}{2} = -2 - \log_n \Psi(n) < 0$, $n^{k(\kappa - \frac{\alpha}{2})} < 1$, followed by using the appropriate Taylor series expansion. Combining these two results, we get the following upper bound on $C_{\rm s_{23}}$
\begin{align}
C_{\rm s_{23}} \leq O\left( n^{1-\frac{\kappa}{\alpha}} \right) + O(\sqrt{n}) = O\left( n^{1-\frac{\kappa}{\alpha}} \right).
\label{eq:CS23_f}
\end{align}
Substituting \eqref{eq:CS21_f}, \eqref{eq:CS22_f} and \eqref{eq:CS23_f} in \eqref{eq:CS2_f_int}, $C_{\rm s_2}$ can be upper bounded as
\begin{align}
C_{\rm s_2} &= C_{\rm s_{21}} - C_{\rm s_{22}} + C_{\rm s_{23}} \nonumber \\
&\leq O\left(n^{1-\frac{\kappa}{\alpha}}\right) + O\left(n^{1-\frac{\kappa}{\alpha} } \log(n)\right) + O\left( n^{1-\frac{\kappa}{\alpha}} \right) \nonumber \\
&= O\left(n^{1-\frac{\kappa}{\alpha} } \log n \right).
\label{eq:CS2_final}
\end{align}
Now combining \eqref{eq:CS1_final} and \eqref{eq:CS2_final}, $C_{\rm s}$ can be upper bounded as
\begin{align}
C_{\rm s} &= C_{\rm s_1} + C_{\rm s_2} = O(n^{1-\frac{\kappa}{\alpha}}) + O\left(n^{1-\frac{\kappa}{\alpha} } \log(n)\right)  \nonumber \\
&= O\left(n^{1-\frac{\kappa}{\alpha} } \log n\right),
\end{align}
which completes this proof.
\hfill \IEEEQED

\subsection{Proof of Lemma~\ref{lem:LB_D}} \label{app:LB_D}
Let the source-destination separation of a randomly chosen pair be $D_{B_n}$. Denote the locations of the source and destination BSs for this pair by $(X_{\rm s},Y_{\rm s}) \in \nbbR^2$ and $(X_{\rm d},Y_{\rm d}) \in \nbbR^2$, where $X_{\rm s}$ and $Y_{\rm s}$ are i.i.d. random variables uniformly distributed in $[0, \sqrt{n}]$. Similarly $X_{\rm d}$ and $Y_{\rm d}$ are i.i.d. and uniformly distributed in $[0, \sqrt{n}]$. The cumulative distribution function (CDF) of $D_{B_n}$ is
\begin{align}
\nbbP(D_{B_n} \leq z) &= \nbbP\left((X_{\rm s} - X_{\rm d})^2 + (Y_{\rm s} - Y_{\rm d})^2 \leq z^2\right) \nonumber\\
&\leq \nbbP\left((X_{\rm s} - X_{\rm d})^2 \leq z^2\right) \nonumber\\
&= \nbbP\left( -z+X_{\rm d} \leq X_{\rm s} \leq z+X_{\rm d} \right) \nonumber\\
&\leq \frac{2z}{\sqrt{n}},
\end{align}
where the inequality in the last step is because we ignored the restrictions on the range of $X_{\rm s}$.  Clearly $\lim_{n\rightarrow \infty} \nbbP(D_{B_n} \leq n^{\frac{1}{2} - \epsilon}) = 0$ for $\epsilon>0$, from which the result follows. \hfill\IEEEQED

\subsection{Proof of Lemma~\ref{lem:sh_uniform}} \label{app:sh_uniform}
First note that the exponential term can be equivalently expressed as
 $\exp\left(-j 2\pi D \Psi(n) \right) =  \exp\left(-j 2\pi X \right)$, 
where $X = D \Psi(n) \mod 1$, which clearly lies in $[0,1]$. It is enough to show that as $\Psi(n)$ grows large, $X$ tends to a uniform distribution between $[0,1]$. Therefore for $x \in [0,1]$
\begin{align}
\nbbP(X\leq x) &= \sum_{i = 0}^{\infty} \nbbP(i \leq D \Psi(n) \leq i + x) \nonumber \\
&= \sum_{i = 0}^{\infty} \nbbP\left(\frac{i}{\Psi(n)} \leq D  \leq \frac{i + x}{\Psi(n)}\right)\\
&\stackrel{(a)}{=} x \sum_{i = 0}^{\infty} \frac{1}{\Psi(n)} f_{D} \left(\frac{i}{\Psi(n)} \right) = x,
\end{align}
where $(a)$ and its subsequent step hold under $\Psi(n) \rightarrow \infty$ and $f_D(\cdot)$ denotes the probability density function of $D$. Clearly $X$ is uniformly distributed in $[0, 1]$. \hfill\IEEEQED

\subsection{Supplement to Section~\ref{sec:ShortHop_Lattice}: Achievability of $R_{\rm sh}(n) = \Omega(\sqrt{n})$ using $\Psi(n) = \sqrt{n}$.}
\label{app:achievability}

Recall that in Section~\ref{sec:ShortHop_Lattice} we showed that $\Psi(n) = \sqrt{n}$ is sufficient to achieve $\nbbE[R_{\rm sh}(n)] = \Omega(\sqrt{n})$.
However, this does not imply scalability w.h.p. for a given realization of the BS antenna locations. 
To address this, we numerically evaluated the distribution of achievable rate $R_{\rm sh}(n)$ in Section~\ref{sec:Discussion} 
and showed that the achievable rate (i) concentrates around its mean, and (ii) is lower bounded w.h.p. by our average rate lower bound. 
In this Appendix, we present a proof idea based on the analysis and insights of \cite{DesLevC2013b} on LoS MIMO channels, where it was shown that the number 
of significant singular values of the LoS MIMO channel matrix is $\frac{a}{\lambda d}$, where $d$ is the transmitter-receiver separation, and they are roughly of the same order.

Using the fact that the rate lower bound in \eqref{eq:Rshn} 
is decreasing with respect to $\nbR$ on the cone of positive semidefinite Hermitian matrices, and that 
$\lambda_{\max}(\nbR)  \nbI \geq \nbR$, where $\lambda_{\rm max} (\cdot)$ is the spectral radius, we have the lower bound
\begin{align}
R_{\rm sh} (n) \geq \log \det \left( \nbI + \frac{P}{\Psi(n)} \frac{1}{\lambda_{\rm max}(\nbR)} \nbH \nbH^{\dag} \right).
\end{align}
From the definition of $\nbR$ given by \eqref{eq:R_main}, we have
\begin{align}
\lambda_{\rm max} (\nbR) \leq 1 + \frac{P}{\Psi(n)} \sum_{i\in \ncalI} \lambda_{\max} (\nbH^{(i)} \nbH^{(i)\dag})
\label{eq:lambdaR}
\end{align}
According to the analysis of \cite{DesLevC2013b},  $\nbH^{(i)} \nbH^{(i)\dag}$ has $\frac{a}{\lambda d^{(i)}}$ significant eigenvalues, 
each with value (in order) $\frac{\left[\min\left\{1,\left(d^{(i)}\right)^{-\frac{\alpha}{2}}\right\} \right]^2 \Psi(n)^2 \lambda d^{(i)}}{a}$, 
where we used the fact that the sum of the eigenvalues is equal to $\Tr\left (\nbH^{(i)} \nbH^{(i)\dag}\right ) = \left[\min\left\{1,\left(d^{(i)}\right)^{-\frac{\alpha}{2}}\right\} \right]^2 \Psi(n)^2$. To ensure full DoF for the desired link, we have $\frac{a}{\lambda d_{\max}} = \Psi(n)$, implying that the eigenvalues of 
$\nbH^{(i)} \nbH^{(i)\dag}$ can be expressed (in order) as $\left[\min\left\{1,\left(d^{(i)}\right)^{-\frac{\alpha}{2}}\right\} \right]^2 \Psi(n) d^{(i)} d_{\rm max}^{-1}$. 
Substituting this in \eqref{eq:lambdaR} in place of $\lambda_{\max} (\nbH^{(i)} \nbH^{(i)\dag})$, we get
\begin{align}
\lambda_{\rm max} (\nbR) \leq 1 + P \sum_{i\in \ncalI} \left[\min\left\{1,\left(d^{(i)}\right)^{-\frac{\alpha}{2}}\right\} \right]^2 d^{(i)} d_{\rm max}^{-1},
\end{align}
which is the same as the constant term of $\nbbE[\nbR]$ in \eqref{eq:R_int_sum}, except that the above expression contains an extra distance 
term $d^{(i)}$. Following the same procedure as for \eqref{eq:R_int_sum}, we can upper bound $\lambda_{\rm max} (\nbR)$ by a constant, but due to 
the presence of extra $d^{(i)}$, this will be valid for $\alpha > 3$. Note that the above arguments are based on the conjecture that all the eigenvalues of $\nbH^{(i)} \nbH^{(i)\dag}$ are ``exactly'' the same (in order) (hence equal to $\lambda_{\max} (\nbH^{(i)} \nbH^{(i)\dag})$), which is motivated by the discussion in \cite{DesLevC2013b}. To rigorize this proof idea, we need to prove this conjecture and show that the element wise approximations made in \cite{DesLevC2013b} imply convergence to the same limit of the asymptotic distribution of the actual LoS MIMO channel matrix. This is a promising area for future work.
\bibliographystyle{IEEEtran}
\bibliography{WirelessBackhaulNetworks_v3.1_2col}
\end{document}